\documentclass[aps,prb,amsmath,amssymb,reprint,superscriptaddress,preprintnumbers,showpacs,intlimits,shortbibliography]{revtex4-2}
\pdfoutput=1
\bibpunct{[}{]}{,}{n}{}{}

\usepackage{bm,latexsym,mathrsfs,enumerate}
\usepackage{xcolor}
\usepackage{mathtools}
\usepackage[breaklinks=true,unicode=true,urlcolor = blue,colorlinks = true,citecolor = blue,linkcolor = blue]{hyperref}
\usepackage{pgf,tikz}
\usepackage{graphicx}
\usepackage[most]{tcolorbox}
\usepackage{multirow}
\usepackage{wasysym}
\usepackage[mathcal]{euscript}
\usepackage{siunitx}
\renewcommand{\vec}[1]{\bm{#1}}
\newcommand{\ts}[1]{\textsc{#1}}
\newcommand{\dd}{\mathrm{d}}

\newcommand{\x}{\mathcal{X}}
\newcommand{\sgn}[1]{\text{sign}(#1)}
\usepackage[normalem]{ulem}
\begin{document}
%
%\preprint{\textcolor[rgb]{0.00,0.50,0.75}{{\texttt{Draft\_v\_1.22}}}}
%
\title{Curvature induced drift and deformation of magnetic skyrmions: comparison of ferro- and antiferromagnetic cases}
\author{Kostiantyn V. Yershov}
\email{yershov@bitp.kiev.ua}
\affiliation{Bogolyubov Institute for Theoretical Physics of the National Academy of Sciences of Ukraine, 03143 Kyiv, Ukraine}
\affiliation{Leibniz-Institut f\"ur Festk\"orper- und Werkstoffforschung, IFW Dresden, 01171 Dresden, Germany}

\author{Attila K{\'a}kay}
\email{a.kakay@hzdr.de}
\affiliation{Helmholtz-Zentrum Dresden - Rossendorf e.V., Institute of Ion Beam Physics and Materials Research, 01328 Dresden, Germany}

\author{Volodymyr P. Kravchuk}
\email{volodymyr.kravchuk@kit.edu}
\affiliation{Bogolyubov Institute for Theoretical Physics of the National Academy of Sciences of Ukraine, 03143 Kyiv, Ukraine}
\affiliation{Institut f\"ur Theoretische Festk\"orperphysik, Karlsruher Institut f\"ur Technologie, 76131 Karlsruhe, Germany}
\date{\today}
%
%%%%%%%%%%%%%%%%%%%%%%%%%%%%%%%%%%%%%%%%%%%%%%%%%%%%%%%%%%%%%%%%%%%%%
%
%         ABSTRACT
%
%%%%%%%%%%%%%%%%%%%%%%%%%%%%%%%%%%%%%%%%%%%%%%%%%%%%%%%%%%%%%%%%%%%%%
%
\begin{abstract}
The influence of the geometrical curvature of chiral magnetic films on the static and dynamic properties of  hosted skyrmions are studied theoretically.  We predict the effects of the curvature-induced drift of skyrmions under the action of the curvature gradients without any external stimuli. The strength of the curvature-induced driving force essentially depends on the skyrmion type, N\'eel or Bloch, while the trajectory of motion is determined by the type of magnetic ordering: ferro- or antiferromagnetic. When moving on the surface, skyrmions undergo deformations that depend on the type of skyrmion. In the small-curvature limit, using the collective-variable approach we show, that the driving force acting on a N{\'e}el skyrmion is linear in the gradient of the mean curvature. The driving acting on a Bloch skyrmion is much smaller: it is proportional to the product of the mean curvature and its gradient. In contrast to the fast N{\'e}el skyrmions, the dynamics of the slow Bloch skyrmions is essentially affected by the skyrmion profile deformation. For the sake of simplicity we restrict ourselves to the case of zero Gaussian curvature and consider cylindrical surfaces of general type. Equations of motion for ferromagnetic and antiferromagnetic skyrmions in curved magnetic films are obtained in terms of collective variables. All analytical predictions are confirmed by numerical simulations.
\end{abstract}
\maketitle
%
%%%%%%%%%%%%%%%%%%%%%%%%%%%%%%%%%%%%%%%%%%%%%%%%%%%%%%%%%%%%%%%%%%%%%
%
%         INTRODUCTION
%
%%%%%%%%%%%%%%%%%%%%%%%%%%%%%%%%%%%%%%%%%%%%%%%%%%%%%%%%%%%%%%%%%%%%%
%
\section{Introduction}
The recently established intimate relation between chiral magnetic interactions and the system geometry \cite{Hill21,Gaididei14} opens up a new direction for the study of curvature-induced properties of magnetic skyrmions, as well as other topological solitons. It was shown that an effective Dzyaloshinsky-Moriya interaction (DMI) appears in the isotropic Heisenberg model considered in a curvilinear space \cite{Hill21,Gaididei14,Sheka15}. Moreover, a curvilinear magnetic film is equivalent to a planar film, where the curvature is replaced by the effective DMI and anisotropy interactions \cite{Pylypovskyi18a,Carvalho-Santos20}. These features make chiral magnetism similar in spirit to the general relativity \cite{Hill21}. Curvature of magnetic films break symmetries related to the topological charge \cite{Elias19,Vojkovic17} and the chirality \cite{Pylypovskyi15b,Kravchuk12a,Sloika14} of topological solitons. Magnetic skyrmions, being a topological soliton in chiral magnets \cite{Bogdanov89r,Wiesendanger16,Fert17}, naturally appeared in the focus of study of curvature effects. On one hand, skyrmions are promising key elements for realization of nonvolatile memory and logic devices \cite{Fert13,Finocchio16,Wiesendanger16,Fert17,Back20}. On the other hand, the three-dimensional generalization of traditional one- and two-dimensional magnetic elements anticipates novel features in the spintronic devices, e.g. miniaturization, increased capacity and speed of the data processing \cite{Fischer20,Fernandez17,Parkin08}. In this situation, the emergence of numerous questions about properties of skyrmions on curved surfaces have stimulated noticeable research activity in the skyrmionic branch of modern magnetism \cite{Sheka21b,Makarov21}. A number of interesting results are already obtained. It was demonstrated that skyrmions can be stabilized on a curvilinear shell without intrinsic DMI \cite{Kravchuk16a,Pylypovskyi18a,Yang21,Yang21a}. Skyrmions can be pinned by a local curvature deffect \cite{Kravchuk18a,Pavlis20}. The pinned skyrmion can demonstrate a multiplet of equilibrium states with different radii and one of these states can be the ground state of the system \cite{Kravchuk18a}. The latter offers the possibility to fabricate zero-field skyrmion lattices of arbitrary symmetry \cite{Kravchuk18a}. Furthermore, on closed cylindrical surfaces, the current-driven skyrmion dynamics is achievable in the high current density regime \cite{Wang19}, thus reaching stable transport with high speeds. This is in contrast to the case of planar stripes, where for high current densities skyrmions annihilate at the boundary due to the skyrmion Hall effect \cite{Jiang16a}.

The present manuscript contributes to the study of curvature effects on magnetic skyrmions. Here, we propose a comprehensive theory of the curvature-induced drift of skyrmions along cylindrical surfaces with curvature gradients. This effect is analogous to the curvature-induced drift of domain walls in curvilinear wires \cite{Yershov18a}. We consider two different types of magnetic ordering, ferromagnetic (FM) and antiferromagnetic (AFM), and two types of skyrmions, N{\'e}el and Bloch. Our analysis is based on the collective variables approach generalized for curvilinear coordinates \cite{Korniienko20} and on micromagnetic simulations.

The general dynamical properties of FM and AFM skyrmions are formulated in Sections~\ref{sec:dynamics-FM} and \ref{sec:dynamics-AFM}, respectively. We show that both types of skyrmions experience a curvature-induced deformation of their profiles, and that the dynamics of Bloch skyrmions is strongly affected by the deformation. The corresponding theory of the curvature induced skyrmion deformation is provided in Section~\ref{sec:stat-skyr}. We illustrate our general results on two examples of cylindrical surfaces with sinusoidal and spiral directrices. The formulation of the model for FM and AFM curvilinear films is provided at the very beginning in Section~\ref{sec:model}.
%The paper starts with formalization of the model of curvilinear film provided in Section~\ref{sec:model}. 
Details of analytical calculations, as well as details of micromagnetic simulations are explained in Appendices~\ref{app:energy}-\ref{app:simuls}. In supplemental materials we provide four movies, which show the curvature induced skyrmion dynamics of N{\'e}el and Bloch skyrmions in different regimes.

\section{Model of a curvilinear magnetic film}\label{sec:model}
In the following we consider two cases of chiral magnets with different types of magnetic ordering: ferro- and antiferromagnetic. The temperature is assumed to be well bellow the ordering point -- Curie and N{\'e}el temperatures for ferro- and antiferromagnets, respectively. In this case, the local order parameter for a FM can be presented by means of a continuous unit-vector field $\vec{\Omega}(\vec{r},t)=\vec{M}/M_s$, where $\vec{M}(\vec{r},t)$ is the magnetization density and $M_s$ is the saturation magnetization. Considering an AFM we restrict ourselves to the fully compensated two-sublattice AFMs in the so-called exchange approximation \cite{Baryakhtar79,Ivanov95e,Turov01en}, i.e. the  exchange field, which keeps the magnetization of the two AFM sublattices $\vec{M}_1$ and $\vec{M}_2$ antiparallel \footnote{The action of the exchange field $B_\ts{x}$ can be expressed in terms of the uniform exchange interaction between two sublattices. The interaction energy density is $\mathscr{E}_\ts{x}^\ts{u} =B_\ts{x}\left(\vec{M}_1\cdot\vec{M}_2\right)/(2M_s)$.}, is much larger than all other effective magnetic fields in the system. In this case the AFM can also be described by a single unit vector field $\vec{\Omega}(\vec{r},t)=(\vec{M}_2-\vec{M}_1)/(2M_s)$, that is the N{\'e}el order parameter. Here, $M_s$ denotes the saturation magnetization of each of the sublattices. In the considered limit, energy of both FM and AFM films can be modeled by means of the same Hamiltonian \cite{Bogdanov02a}
\begin{equation}\label{eq:model}
	E[\vec{\Omega}]=L\int_{\mathcal{S}}\left[A\mathscr{E}_\textsc{x}+D\mathscr{E}_\textsc{d}+K\left(1-\Omega_n^2\right)\right]\textrm{d}\mathcal{S},
\end{equation}
where the integral is taken over the area of the film surface and we assumed that the film thickness $L$ is small enough to ensure a uniform magnetization in the direction of the film normal $\vec{n}$. The first term in \eqref{eq:model} is the energy density of the nonuniform exchange $\mathscr{E}_\textsc{x}=\sum_{i=x,y,z}\left(\partial_i\vec{\Omega}\right)^2$, where $A$ is the exchange stiffness. The second term in~\eqref{eq:model} corresponds to the DMI, with $D$ being the DMI constant. In the following we consider two types of DMI, namely the interfacial $\mathscr{E}_\textsc{d}^\textsc{n}=\Omega_n\vec{\nabla}\cdot\vec{\Omega}-\vec{\Omega}\cdot\vec{\nabla}\Omega_n$ and isotropic $\mathscr{E}_\textsc{d}^\textsc{b}=\vec{\Omega}\cdot\left[\vec{\nabla}\times\vec{\Omega}\right]$ types  which support the formation of skyrmions of N{\'e}el and Bloch types, respectively. Here $\Omega_n=\vec{n}\cdot\vec{\Omega}$. The last term in~\eqref{eq:model} represents the uniaxial anisotropy with $K>0$ being the easy-normal anisotropy constant.

\begin{figure*}
	\includegraphics[width=\textwidth]{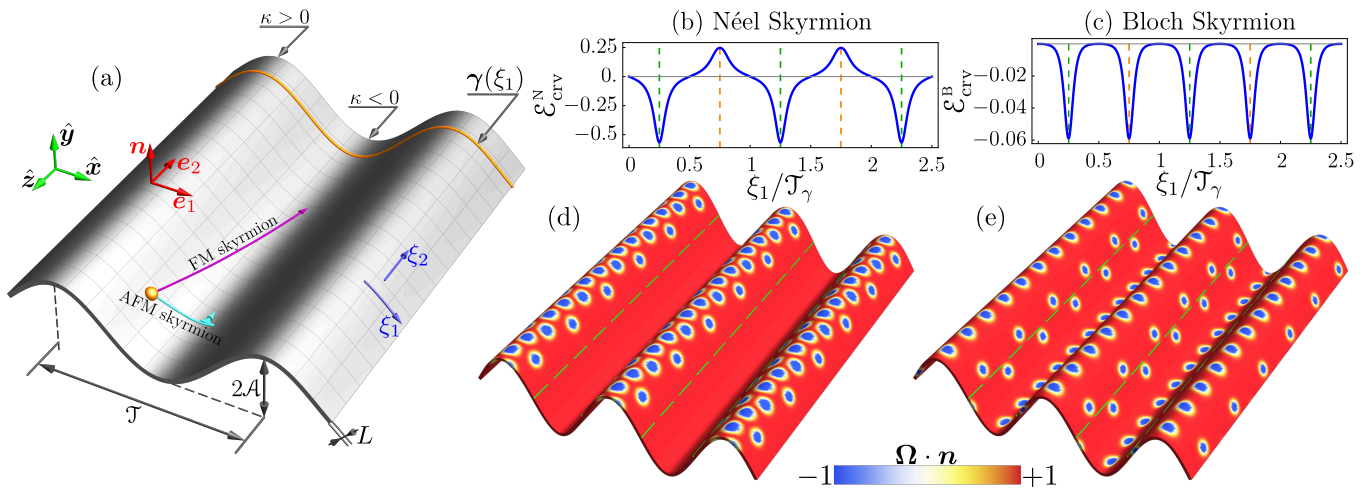}
	\caption{\label{fig:sine_geometry}%
		(Color online) (a) The cylindrical film with sinusoidal directrix \eqref{eq:sin} shown by the orange line.  Magenta and cyan lines demonstrate possible trajectories of FM and AFM Bloch skyrmions discussed in Sections \ref{sec:dynamics-FM} and \ref{sec:dynamics-AFM}, respectively. Panels (b) and (c) show the variation of energies of N{\'e}el $\mathcal{E}^\textsc{n}_\text{crv}$ and Bloch $\mathcal{E}^\textsc{b}_\text{crv}$ skyrmions along the directrix (direction $\xi_1$) obtained by means of \eqref{eq:ser-N} and \eqref{eq:ser-B}, respectively. $\mathcal{T}_{\gamma}$ is the period of the directrix in terms of the arc length. Green and orange vertical dashed lines correspond to the position of the convexities and concavities, respectively. (d) and (e) -- quasi-equilibrium distribution of FM N{\'e}el and Bloch skyrmions on the curved film, respectively. Green dashed lines in (d) and (e) correspond to the curvature $\kappa = 0$. Data in (d) and (e) is obtained by means numerical simulations for $\mathcal{A}=10\ell$, $\mathcal{T}=35\ell$, (corresponds to $\mathcal{T}_\gamma\approx55.1\ell$), and $d=1$, $\eta_{\textsc{g}}=0.1$. Skyrmion dynamics preceding the presented states is shown in Supplemental movies \cite{Note2}. Total simulation time in both cases is $t_\text{sim} = 10^4/\omega_{\textsc{fm}}$.}
\end{figure*}

For the sake of simplicity we consider generalized cylindrical surfaces as cases of study. 
Let the cylinder generatrix is directed along $\hat{\vec{z}}$ and directrix $\vec{\gamma}$ lies in a perpendicular plane $x0y$. The cylinder surface is determined by  the parameterization $\vec{\varsigma}(x_1,x_2) = \vec{\gamma}(x_1)-x_2\hat{\vec{z}}$, where $\hat{\vec{z}}\cdot\partial_{x_1}\vec{\gamma}=0$, see Fig.~\ref{fig:sine_geometry}a.
Let coordinate $x_1$ be the arclength of the curve $\vec{\gamma}$ and coordinate $x_2=-z$. The parameterization $\vec{\varsigma}(x_1,x_2)$ induces Eucledian metric $g_{\alpha\beta}=\delta_{\alpha\beta}$ on the surface and the tangential basis $\vec{e}_\alpha = \partial_\alpha\vec{\varsigma}$, where $\vec{e}_1=\partial_{x_1}\vec{\gamma}$ is the unit vector tangential to $\vec{\gamma}$ and $\vec{e}_2=-\hat{\vec{z}}$. The unit surface normal we introduce as $\vec{n}=\vec{e}_1\times\vec{e}_2$. Curvilinear properties of the surface $\vec{\varsigma}$ are completely determined by the only parameter -- the signed curvature $\kappa(x_1)$ of the curve $\vec{\gamma}$. We define the sign of $\kappa$ by means of the convention $\partial_{x_1}^2\vec{\gamma}=-\kappa\vec{n}$, i.e. $\kappa>0$ and $\kappa<0$ for convexities and concavities, respectively, see Fig.~\ref{fig:sine_geometry}a.

\section{Static skyrmions in cylindrical surfaces}\label{sec:stat-skyr}
Our aim is to estimate the skyrmion energy in the limit of small curvature
\begin{equation}\label{eq:limit}
	1\gg|\kappa|r_0\gg|\kappa'|r_0^2\gg|\kappa''|r_0^3\dots
\end{equation}
where prime denotes the derivative with respect to $x_1$, and $r_0$ is skyrmion radius. To this end we introduce the angular parameterization
\begin{equation}\label{eq:ang-par}
\vec{\Omega}=\sin\theta\vec{\varepsilon}+\cos\theta\vec{n},\qquad \vec{\varepsilon}=\cos\phi\vec{e}_1+\sin\phi\vec{e}_2\end{equation}
for the order parameter and consider two successive approximations of the curvature effect. 

\subsection{Rigid skyrmion: the first order perturbation in curvature.}

Let us first neglect possible deformations of the skyrmion profile caused by the curvature. We consider the curvature as a small perturbation which does not change the skyrmion profile, however it can shift the skyrmion energy. In other words, we are looking for the energy corrections induced by the curvature in the first order perturbation theory. Taking the advantage of Euclidean metric we apply the following Ansatz for the skyrmion structure

\begin{equation}\label{eq:Ansatz-rigid}
	\begin{split}
		&\theta=\theta_0(r),\qquad r=\sqrt{(x_1-X_1)^2+(x_2-X_2)^2} ,\\
		&\phi=\chi+\Phi_0,\qquad \chi=\arctan\frac{x_2-X_2}{x_1-X_1}	
	\end{split}
\end{equation}
which structurally coincides with the planar skyrmion solution. Here, $\theta_0(r)$ represents the profile of the planar skyrmion and $X_1$, $X_2$ are curvilinear coordinates of the skyrmion center on the surface. The constant $\Phi_0$ depends on the skyrmion type: $\Phi_0=\Phi_0^{\ts{n}}=(1-\sigma)\pi/2$ for N{\'e}el and $\Phi_0=\Phi_0^{\ts{b}}=\sigma\pi/2$ for Bloch skyrmions. $\sigma=\sgn{D}=\pm1$, and the boundary conditions assumed for the skyrmion profile are $\theta_0(0)=\pi$, $\theta_0(\infty)=0$.

As a next step we calculate the energy of the skyrmion on the curved film. To this end we substitute Ansatz \eqref{eq:Ansatz-rigid} into the Hamiltonian \eqref{eq:model} and perform integration over the film area. Note, that the energy densities $\mathscr{E}_\textsc{x}$ and $\mathscr{E}_\textsc{d}$ depend on curvature $\kappa(x_1)$, which in the limit \eqref{eq:limit} can be presented in form of series
\begin{equation}\label{eq:kappa-ser}
	\kappa(x_1)=\kappa(X_1)+\kappa'(X_1)r\cos\chi+\frac{\kappa''(X_1)}{2}(r\cos\chi)^2+\dots
\end{equation}
The polar coordinates $(r,\chi)$ used here are defined in \eqref{eq:Ansatz-rigid}. Using \eqref{eq:kappa-ser} under the integral \eqref{eq:model} we obtain the energy of the skyrmion as a function of its position $X_1$ (for details see Appendix~\ref{app:energy}). The normalized total energy $\mathcal{E}^{\textsc{n},\textsc{b}}=E^{\textsc{n},\textsc{b}}/(8\pi AL)$ of N{\'e}el ($\mathcal{E}^{\textsc{n}}$), as well as Bloch ($\mathcal{E}^{\textsc{b}}$) skyrmions, can be presented as a sum $\mathcal{E}^{\textsc{n},\textsc{b}}=\mathcal{E}^{\textsc{n},\textsc{b}}_{\text{pl}}+\mathcal{E}^{\textsc{n},\textsc{b}}_{\text{crv}}+\text{const}$ of the skyrmion energy in planar film $\mathcal{E}^{\textsc{n}}_{\text{pl}}=\mathcal{E}^{\textsc{b}}_{\text{pl}}=\mathcal{C}_0-\mathcal{C}_2$ and the curvature-induced corrections
\begin{subequations}\label{eq:Ecrv}
	\begin{align}
		\label{eq:ser-N}&\mathcal{E}^{\textsc{n}}_{\text{crv}}=\mathfrak{C}^{\textsc{n}}_1\varkappa+\mathfrak{C}^{\textsc{n}}_2\frac{\varkappa^2}{2}+\mathcal{O}(\varkappa'',{\varkappa'}^2),\\
		\label{eq:ser-B}&\mathcal{E}^{\textsc{b}}_{\text{crv}}=\mathfrak{C}^{\textsc{b}}_2\frac{\varkappa^2}{2}+\mathcal{O}(\varkappa'^2).
	\end{align}
\end{subequations}
Here, $\varkappa(\mathcal{X}_1)=\kappa(X_1)\ell$ is the dimensionless curvature at the point of the skyrmion location $\mathcal{X}_1=X_1/\ell$. The magnetic length is $\ell=\sqrt{A/K}$. As seen from equations~\eqref{eq:Ecrv} the total skyrmion energy depends on the skyrmion position. Constants $\mathcal{C}_0=\frac14\int_{0}^\infty[\theta_0'(\rho)^2+\rho^{-2}\sin^2\theta_0(\rho)]\rho\,\dd \rho$ and $\mathcal{C}_2=\frac14\int_{0}^\infty\sin^2\theta_0(\rho)\rho\,\dd\rho$ are determined by the profile of skyrmions in a planar film, where $\rho=r/\ell$ is the dimensional distance from the skyrmion center. Coefficients in the expansions \eqref{eq:Ecrv} are  $\mathfrak{C}^{\textsc{n}}_1=-\mathcal{C}_2(2+d^2)/d$ and $\mathfrak{C}^{\textsc{n}}_2=\mathfrak{C}^{\textsc{b}}_2=-\mathcal{C}_2$ with $d=D/\sqrt{AK}$ being the dimensionless DMI constant. Note, that $|d|<4/\pi$ is the only control parameter of the model \eqref{eq:model}. Details of the derivation of expansion \eqref{eq:Ecrv} and its higher order terms can be found in Appendix~\ref{app:energy}.

Equations~\eqref{eq:Ecrv} show the principal difference between N{\'e}el and Bloch skyrmions in the curved films, namely the leading terms in expansions \eqref{eq:Ecrv} are linear and quadratic in curvature for the N{\'e}el and Bloch skyrmions, respectively. Since the curvature is small, N{\'e}el skyrmions experience a much stronger influence by the curvature compared to Bloch skyrmions. The reason is that the curvature-induced effective DMI originating from the isotropic exchange interaction is of the interfacial DMI (N{\'e}el) type \cite{Kravchuk16a,Kravchuk18a}. So, for the case of N{\'e}el skyrmions, a direct competition between the intrinsic and the curvature induced DMIs takes place. This makes N{\'e}el skyrmions more sensitive to the film curvature as compared to Bloch skyrmions.

Since $\mathcal{C}_2>0$, the sign of the coefficient $\mathfrak{C}^{\textsc{n}}_1$ at the leading term in \eqref{eq:ser-N} coincides with the sign of the DMI constant. This enables us to expect a drift of N{\'e}el skyrmions towards maxima (minima) of the curvature for $D>0$ ($D<0$). From \eqref{eq:ser-N} we conclude that the position of stable equilibrium $\mathcal{X}_1^0$ for a \emph{N{\'e}el skyrmion} is determined by the condition
\begin{equation}\label{eq:stbl-eqlbr-N}
	\left.\varkappa'\right|_{\x_1^0}=0,\qquad \left.\sigma\varkappa''\right|_{\x_1^0}<0.
\end{equation}
Note, that $\x_1=\x_1^0$ determines a straight line  parallel to $\hat{\vec{z}}$ on the cylindrical surface.

Regardless of the DMI sign, the coefficient $\mathfrak{C}^{\textsc{b}}_2$ at the leading term in \eqref{eq:ser-B} is negative. This means that the Bloch skyrmion drifts towards curvature extremes. However, the driving force is expected to be much smaller as for the N{\'e}el skyrmion. From \eqref{eq:ser-B} we conclude that the position of the stable equilibrium $\mathcal{X}_1^0$ for a \emph{Bloch skyrmion} is determined by the condition
\begin{equation}\label{eq:stbl-eqlbr-B}
	\left.\varkappa\varkappa'\right|_{\x_1^0}=0,\qquad \left.\varkappa'^2+\varkappa\varkappa''\right|_{\x_1^0}<0.
\end{equation}
It follows from \eqref{eq:stbl-eqlbr-B} that the line $\x_1=\x_1^0$ determined by $\varkappa(\x_1^0)=0$ always corresponds to an unstable equilibrium.

We verify the predictions \eqref{eq:stbl-eqlbr-N} and \eqref{eq:stbl-eqlbr-B} about different equilibrium positions for N{\'e}el and Bloch skyrmions by means of micromagnetic simulations performed for a sinusoidal cylindrical surface with directix
\begin{equation}\label{eq:sin}
\vec{\gamma}=x\hat{\vec{x}}+\mathcal{A}\sin(qx)\hat{\vec{y}}.
\end{equation} 
Using the expression for the directix curvature $\kappa=\mathcal{A}q^2\sin(qx)/[1+\mathcal{A}^2q^2\cos^2(qx)]$ and conditions \eqref{eq:stbl-eqlbr-N} and \eqref{eq:stbl-eqlbr-B} we obtain that the positions of stable equilibrium for N{\'e}el and Bloch skyrmions correspond to $x_0=(\frac14+n)\mathcal{T}$ and $x_0=(\frac14+\frac{n}{2})\mathcal{T}$, respectively. Here $\mathcal{T}=2\pi/q$ is period of the surface and it was assumed that $D>0$. As an initial state for the simulations we choose a diluted skyrmion lattice with the uniform skyrmion distribution along the surface. The numerical time integration of the Landau-Lifshitz equation leads to skyrmion dynamics which results in the essentially nonuniform final skyrmions distribution shown in Figs.~\ref{fig:sine_geometry}d,e. See also the supplemental movies \footnote{Reference to Supplemental materials}. Namely, N{\'e}el skyrmions are concentrated near convexities of the surface, while Bloch skyrmions are pinned at convexities as well as at concavities. This is in complete agreement with the equilibrium stationary positions obtained from \eqref{eq:stbl-eqlbr-N} and \eqref{eq:stbl-eqlbr-B}, and it is also consistent with the recent prediction about pinning of the Bloch skyrmion on the top of the Gau{\ss}ian shaped cylindrical defect of a planar stripe \cite{Carvalho-Santos21}. For details of micromagnetic simualtions see Appendix~\ref{app:simuls}. Note that the curvature-induced potential $\mathcal{E}^{\textsc{b}}_{\text{crv}}$ for Bloch skyrmions is very flat in the vicinity of the unstable equilibrium line $\kappa=0$, see Fig.~\ref{fig:sine_geometry}c. Due to the vanishing driving force the Bloch skyrmions in this region move extremely slow. Although, Bloch skyrmions do not leave the vicinity of unstable equilibrium during simulation time $t_{\text{sim}}=10^4/\omega_{\textsc{fm}}$ (Fig.~\ref{fig:sine_geometry}e), the supplemental video \cite{Note2} shows that they move away from the line $\kappa=0$. Here $\omega_{\textsc{fm}}$ is frequency of the uniform ferromagnetic resonance in the planar film, see Sec.~\ref{sec:dynamics-FM} for the explicit definition.
%In this region, the skyrmion dynamics can be influenced by additional small potential deformations which arise due to the effects neglected in the model, e.g. pinning potentials arising from the mesh discreteness, repulsion from the skyrmions already pinned in the equilibrium positions. For this reason, in the simulations, we obtain a number of Bloch skyrmions pinned in the region of flatness of the potential $\mathcal{E}^{\textsc{b}}_{\text{crv}}$, see Fig.~\ref{fig:sine_geometry}e.

Described above effect of skyrmion pinning on the curvature extremes is a two-dimensional generalization of the previously described effect of pinning of domain wall on a local bend of the wire \cite{Yershov15b}.

Although the model of rigid skyrmion enables us to find positions of stable equilibrium, the noticeable radial and elliptical deformations of skyrmions were found in simulations. In the following subsection we study this effect in details.

\subsection{Deformed skyrmion: the second order perturbation in curvature.}

\begin{figure*}
	\includegraphics[width=\textwidth]{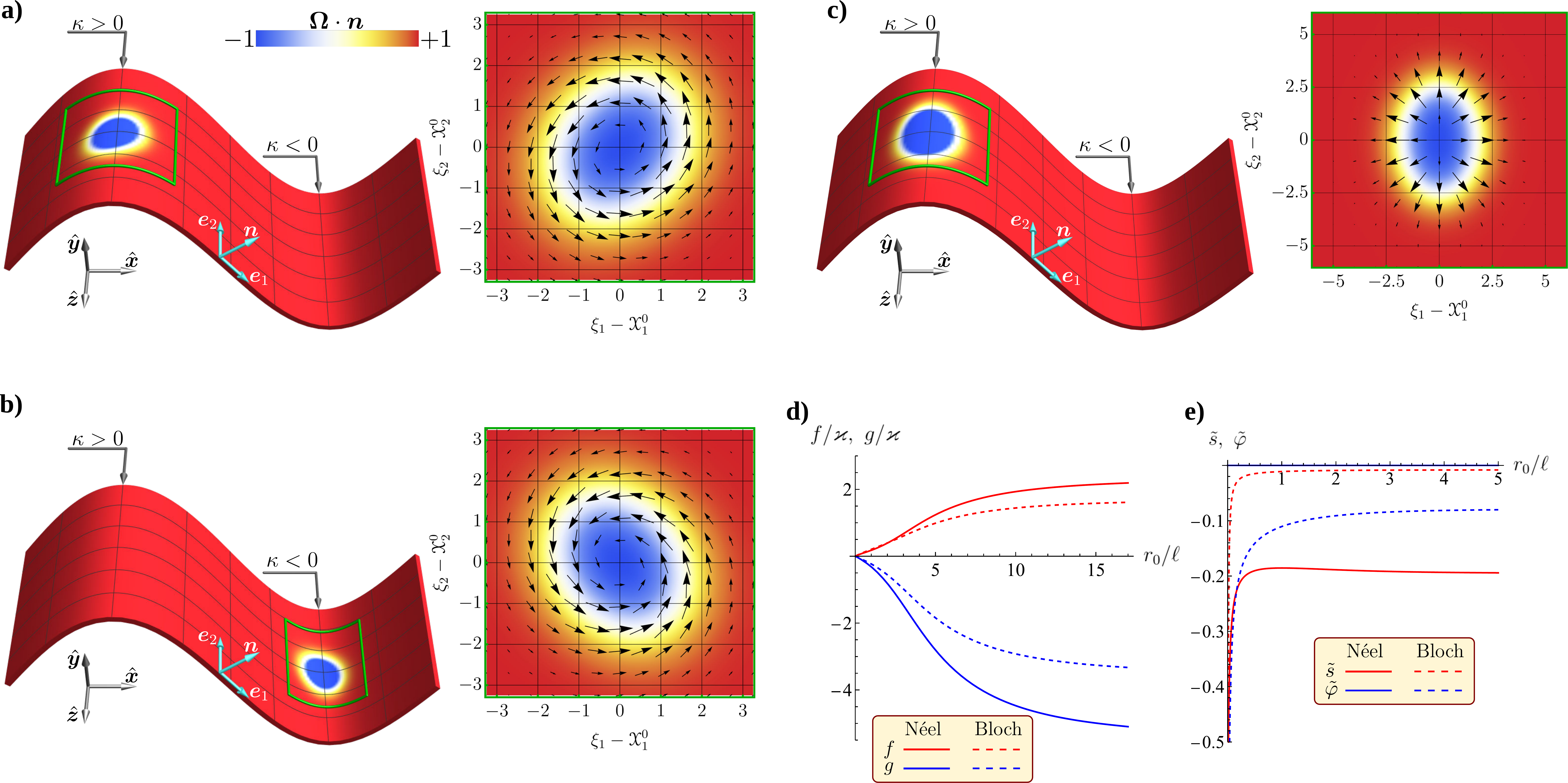}
	\caption{\label{fig:deformations}%
		(Color online) Deformations of Bloch skyrmions in stable equilibrium positions with positive and negative curvatures are shown in panels a) and b), respectively. Deformation of N{\'e}el skyrmion in the position of stable equilibrium is shown in panel c). Data are obtained by means of micromagnetic simulations for the case of FM, geometrical parameters of the surface are the same as in Fig.~\ref{fig:sine_geometry}, DMI constant $d=1$ and $d=0.9$ for Bloch and N{\'e}el skyrmions respectively. Deformations and equilibrium positions of AFM and FM skyrmions are the same. Amplitudes of elliptical deformation are obtained from \eqref{eq:cv-ell-N} and \eqref{eq:cv-ell-B} and shown on panel d) as functions of skyrmion radius. Amplitudes of radially-symmetrical deformation are obtained from \eqref{eq:deform-rs-N} and \eqref{eq:deform-rs-B} and shown on panel e) as functions of skyrmion radius for $\varkappa=0.1$.}
\end{figure*} 

In order to take into account possible skyrmion deformations we consider the following modification of the Ansatz \eqref{eq:Ansatz-rigid} 
\begin{equation}\label{eq:Ansatz-ell-rad}
	\begin{split}
		&\theta=\theta_0\left(\rho\left[1+\tilde{s}+f\cos(2\chi+\lambda)\right]\right),\\
		&\phi=\chi+\Phi_0+\tilde{\varphi}+g\sin(2\chi+\lambda).
	\end{split}
\end{equation} 
Here five additional collective variables are introduced. Variables $\tilde{s}$ and $\tilde{\varphi}$ describe the out-of-surface and in-surface components of the radial symmetrical skyrmion deformation, respectively. While $f$ and $g$ are out-of-surface and in-surface amplitudes of the elliptical deformation, respectively. Variable $\lambda$ controls orientation of the elliptically deformed skyrmion core. Here we utilized the fact that the in-plane and out-of-plane components of the skyrmion excitations have the phase shift $\pi/2$ \cite{Kravchuk18,Kravchuk19a}. On the next step we substitute the Ansatz \eqref{eq:Ansatz-ell-rad} into the total energy expression \eqref{eq:model} and minimize it with respect to the five additional collective variables assuming that the skyrmion position is fixed. I.e. that the shape variables $\tilde{s}$, $\tilde{\varphi}$, $f$, $g$ and $\lambda$ are fast as compared to the position variables $(X_1,\,X_2)$. This assumption is supported by the recent finding \cite{Korniienko20} that the translation mode of the skyrmion pinned on the curvilinear defect has much smaller frequency as compared to the shape modes, e.g. breathing and elliptical modes. Mathematical details of this analysis can be found in Appendix~\ref{app:deform}.

\emph{For N{\'e}el skyrmion}, the equilibrium amplitudes of the radial-symmetrical deformation are $\tilde{\varphi}^\textsc{n}=0$, and $\tilde{s}^\textsc{n}\approx-\varkappa\left(d+d^{-1}\right)$ in the linear in $\varkappa$ approximation. Here we also assumed that $|\varkappa/d|\ll1$. Note that if $|\varkappa|$ is comparable or large than $|d|$, then N{\'e}el skyrmion can collapse and switch its helicity \cite{Kravchuk16a}. The latter is consistent with the divergent behavior of the parameter $\tilde{s}^\textsc{n}$ for vanishing $d$. Positive (negative) sign of parameter $\tilde{s}^\textsc{n}$ corresponds to  decrease (increase) of the skyrmion radius. So, if $d>0$ than the skyrmion radius increases for a skyrmion located on a convexity ($\varkappa>0$) and decreases for a skyrmion located on a concavity ($\varkappa<0$). This is in full agreement with the previous results obtained for N{\'e}el skyrmions on spherical shells \cite{Kravchuk16a} and Gau{\ss}ian bumps \cite{Korniienko20}. The vanishing amplitude of the in-surface deformation $\tilde{\varphi}^\textsc{n}$ means that the skyrmion helicity is not affected, i.e. skyrmion stays of purely N{\'e}el type. This also agrees with the previous results \cite{Kravchuk16a,Kravchuk18a,Korniienko20}.

For the elliptical deformation we obtain $\lambda^\textsc{n}=0$ and 
\begin{equation}\label{eq:cv-ell-N}
			f^\textsc{n}\approx\frac{\varkappa}{d}\frac{\alpha_g+2}{\alpha_f\alpha_g-4},\qquad g^\textsc{n}\approx-\frac{\varkappa}{d}\frac{\alpha_f+2}{\alpha_f\alpha_g-4},
\end{equation}
where $\alpha_f=1+\mathcal{C}_2^{-1}\int_0^\infty\theta_0'(\rho)^2\rho\dd \rho$ and $\alpha_g=1+\mathcal{C}_2^{-1}\int_0^\infty\sin^2\theta_0(\rho)\rho^{-1}\dd \rho$. For details of the derivation see Appendix~\ref{app:deform}. The deformation amplitudes are shown in Fig.~\ref{fig:deformations}d by solid lines. For model \eqref{eq:model}, there is a one-to-one correspondence between parameter $d$ and radius $r_0$ of skyrmion on planar film \cite{Kravchuk18,Komineas20a}. Since skyrmion radius is an easily observable in experiment parameter, here and below we plot the quantity of interest as a function of $r_0$. Skyrmion radius is defined as $\theta_0(r_0)=\pi/2$. Fig.~\ref{fig:deformations}d shows that $\sgn{f^\textsc{n}}=\sgn{\varkappa}$, so basing on \eqref{eq:Ansatz-ell-rad} we conclude that the long axis of the elliptically deformed skyrmion core is directed along $\vec{e}_2$ for convexities and along $\vec{e}_1$ for concavities \footnote{Note that angle $\chi$ is measured from direction $\vec{e}_1$.}. This agrees with our micromagnetic simulations, see Fig.~\ref{fig:deformations}c. Note that according to \eqref{eq:stbl-eqlbr-N}, concavity extremum of a cylindrical surfaces is a position of unstable equilibrium for N{\'e}el skyrmion. For this reason we were able to relax N{\'e}el skyrmions in our simulations only for convexities.

\emph{For Bloch skyrmion}, the equilibrium amplitudes of the radial-symmetrical deformation are $\tilde{\varphi}^\textsc{b}\approx-\varkappa/d$, and $\tilde{s}^\textsc{n}\approx0$ in the linear in $\varkappa$ approximation and under assumption $|\varkappa/d|\ll1$. So, the radial symmetrical part of the Bloch skyrmion deformation is opposite as for the N{\'e}el skyrmion, namely deformation of the out-of-surface component vanishes while the skyrmion helicity is strongly affected. The latter effect can be interpreted as a result of competition of the intrinsic DMI of Bloch type and the effective curvature-induced DMI of N{\'e}el type resulting in skyrmion of an intermediate type.

For the elliptical deformation we obtain $\lambda^\textsc{b}\approx\pi/2$ and 
\begin{equation}\label{eq:cv-ell-B}
	\begin{split}
	&f^\textsc{b}\approx\frac{\varkappa d}{\alpha_f\alpha_g-4}\left[\frac{\alpha_g}{2}\left(1+\frac{2}{d^2}\right)-1+\frac{2}{d^2}\right],\\
	&g^\textsc{b}\approx\frac{\varkappa d}{\alpha_f\alpha_g-4}\left[\frac{\alpha_f}{2}\left(1-\frac{2}{d^2}\right)-1-\frac{2}{d^2}\right].
\end{split}
\end{equation}
This means that the elliptically shaped skyrmion core makes angle $\pi/4$ with the directions $\vec{e}_1$ and $\vec{e}_2$.
Corresponding values of the deformation amplitudes are shown in Fig.~\ref{fig:deformations}d by dashed lines. Details of the derivation are presented in Appendix~\ref{app:deform}. %Note, that the found equlibrium orientations of the skyrmion cores of different types are consistent with the orientation of the periodically modulated states on the chiral cylindrical surfaces for the corresponding types of DMI \cite{Yershov20}. 
The analogous elliptical deformation of skyrmion cores was found in micromagnetic simulations for skyrmions on the surface of circular cylinder \cite{Kechrakos20,Wang19}. 

\begin{figure}
	\includegraphics[width=0.9\columnwidth]{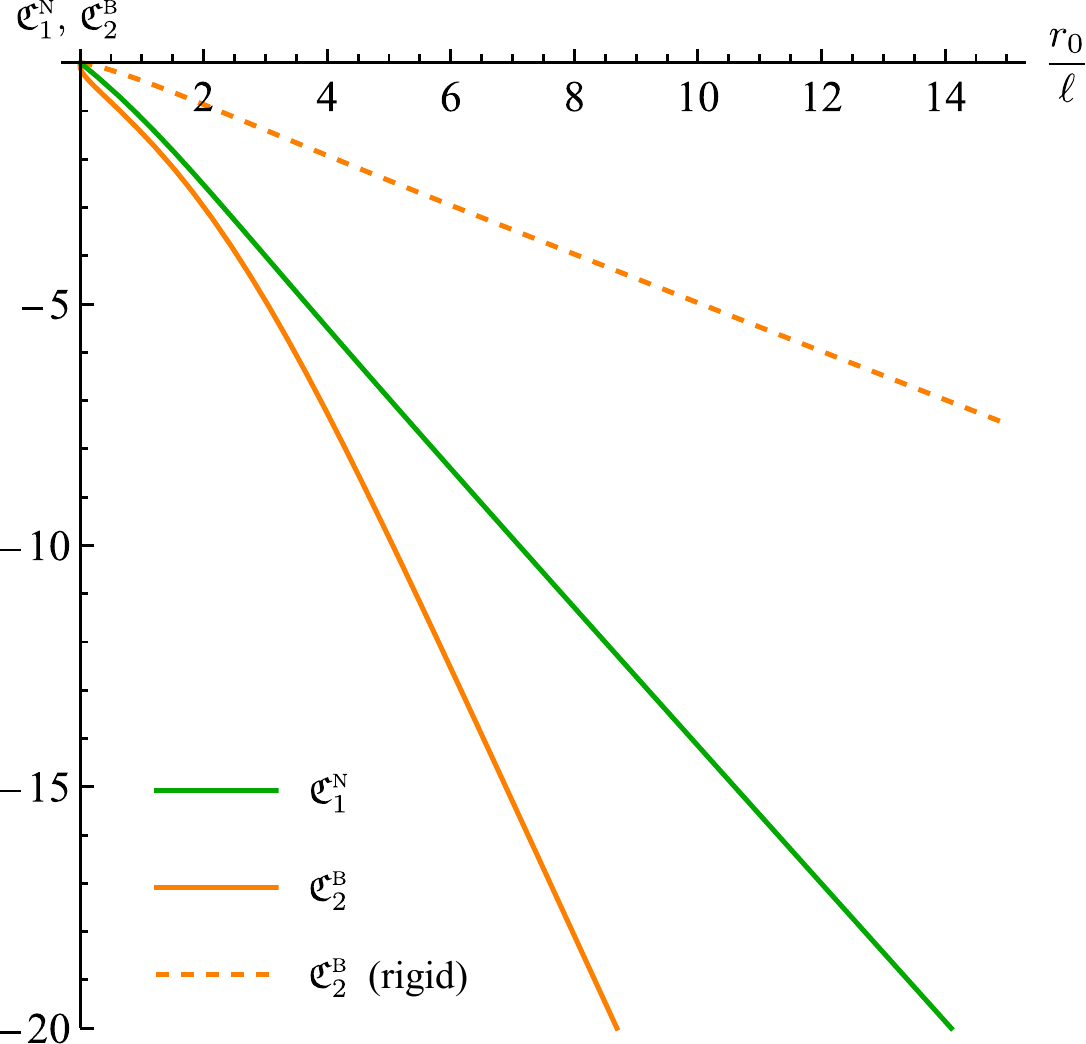}\caption{(Color online) The comparison of the leading terms coefficients in \eqref{eq:Ecrv}: $\mathfrak{C}^{\textsc{n}}_1=-\mathcal{C}_2(2+d^2)/d$, coefficient $\mathfrak{C}^{\textsc{b}}_2=-\mathcal{C}_2(1+\Delta c^\ts{b})$ is presented for $\Delta c^\ts{b}$ determined by \eqref{eq:E-ell-rad-B} (solid line), and for $\Delta c^\ts{b}=0$ (dashed line). The latter case corresponds to rigid skyrmion, when the deformation effects are neglected.  }\label{fig:Cn-Cb}
\end{figure}

On the next step we substitute the obtained above equilibrium values of deformation amplitudes into the energy expression and derive the energy of the deformed skyrmion, see Appendix~\ref{app:deform} for details. The resulting energy structurally coincides with \eqref{eq:Ecrv}, but coefficients $\mathfrak{C}^{\textsc{n},\textsc{b}}_2$ at the terms quadratic in curvature. Coefficient $\mathfrak{C}^{\textsc{n}}_1$ at the linear term remains unchanged, it means that in the limit of small curvature, one can neglect the deformation effects of N{\'e}el skyrmion. This is in strong contrast to the case of Bloch skyrmion, whose energy is essentially affected by the curvature induced deformation. Coefficient $\mathfrak{C}^{\textsc{b}}_2=-\mathcal{C}_2(1+\Delta c^\ts{b})$ obtains the deformation-induced correction
\begin{equation}\label{eq:E-ell-rad-B} 
	\Delta c^\ts{b}=\frac{\left(\dfrac{\mathcal{C}_0}{\mathcal{C}_2}-\dfrac{1}{2}\right)\left(\dfrac{4}{d^2}+d^2\right)\!+\!\alpha_g\!-\!\alpha_f+2\dfrac{\alpha_f\alpha_g}{d^2}}{\alpha_f\alpha_g-4},
\end{equation}
for details see Appendix~\ref{sec:rs-ell-deform}.
The comparison of the leading terms coefficients in \eqref{eq:Ecrv} is presented in Fig.~\ref{fig:Cn-Cb}, which illustrates strong impact of deformation on coefficient $\mathfrak{C}^{\textsc{b}}_2$.

For both types of skyrmions the deformations of different types are independent if the deformation amplitudes are found in approximation linear in $\varkappa$. I.e. the elliptical deformation is not influenced by the radially-symmetrical one and vice-versa; and the corresponding corrections to the energy coefficients are additive, e.g. $\Delta c^\ts{b}=\Delta c^\ts{b}_{\text{rs}}+\Delta c^\ts{b}_{\text{el}}$, where the contributions $\Delta c^\ts{b}_{\text{rs}}$ and $\Delta c^\ts{b}_{\text{el}}$ come from the radially-symmetrical and elliptical deformations, respectively. For the proof see Appendix~\ref{app:deform}.
Above, we considered the case when both types of deformation take place in the same time. However, it is known \cite{Kravchuk18,Kravchuk19a} that for small radius skyrmions the elliptical mode is much more energetic as compared to the radially symmetrical mode. This is correct for both FM and AFM cases, see Fig.~2 in Ref.~\onlinecite{Kravchuk18} and Fig.~6 in Ref.~\onlinecite{Kravchuk19a}. This leads us to the assumption that the elliptical deformation can be neglected for small-radius skyrmions. This assumption we explicitly verify in Appendix~\ref{sec:def-comparison}. The neglect of the elliptical deformation significantly reduces the number of collective variables, from five to two. This enables us to build the nonlinear (in $\varkappa$) theory of the radially-symmetrical deformation and find the deformation amplitudes beyond the restriction $|\varkappa/d|\ll1$. For N{\'e}el slyrmion we obtain
\begin{equation}\label{eq:deform-rs-N}
	\cos\tilde{\varphi}^{\ts{n}}=\sgn{1+\varkappa/d},\quad \tilde{s}^{\ts{n}}=\frac{1-\varkappa d-\frac{\varkappa^2}{2}}{|1+\varkappa/d|}-1,
\end{equation}
see Appendix~\ref{sec:rs}. Note that the case $\varkappa=-d$ corresponds to the complete compensation of the intrinsic DMI by the effective curvature-induced one. In this point skyrmion collapses and changes its helicity afterwards. This is reflected by the jump of $\tilde{\varphi}^{\ts{n}}$ from 0 to $\pi$, and by divergence of the scaling parameter $\tilde{s}^{\ts{n}}$. The effect of the curvature induced helicity switching mediated by the skyrmion collapse was previously described for N{\'e}el skyrmions on spherical shells \cite{Kravchuk16a}. 

In the same manner we obtain
\begin{equation}\label{eq:deform-rs-B}
	\tan\tilde{\varphi}^{\ts{b}}=-\frac{\varkappa}{d},\quad \tilde{s}^{\ts{b}}=\frac{1-\frac{\varkappa^2}{2}}{\sqrt{1+\varkappa^2/d^2}}-1
\end{equation} 
for a Bloch skyrmion, see Appendix~\ref{sec:rs}. Note that for the vanishing intrinsic DMI $d\to0$ one obtains $\tilde{\varphi}^{\ts{b}}\to\pm\pi/2$, i.e. the Bloch skyrmion is transformed into the N{\'e}el one. This agrees with the previous observation that the curvature-induced effective DMI is of interfacial (N{\'e}el) type.

\section{Curvature induced dynamics of FM skyrmions} \label{sec:dynamics-FM}
Since the skyrmion energy on curvilinear film is coordinate dependent, the skyrmion experiences a curvature-induced driving force. The latter results in drift of the skyrmion along the surface, analogous to the drift of a domain wall in a wire under the action of gradients of the wire curvature \cite{Yershov18a}. The aim of this section is to provide an analytical description of the curvature-induced skyrmion drift. %We base ourselves  on the results of the previous section, where we obtained skyrmion energy as a function of its position on the surface, see Eq.~\eqref{eq:Ecrv}. 
The following analysis is based on the assumption that two dynamical processes of the change of skyrmion position and change of its form have different time scales which can be well separated. Namely, we assume that the typical equilibration time of the skyrmion deformations is much smaller as compared to the typical times of the skyrmion displacements. This assumption is supported by the fact that the translation mode of skyrmion on a curvilinear film is much less energetic as compared to the shape modes \cite{Korniienko20}.

Dynamics of a ferromagnetic media is governed by Landau-Lifshitz-Gilbert equation \cite{Landau35,Gilbert04}
\begin{equation}\label{eq:LLG}
	\partial_t \vec{\Omega} = \frac{\gamma_0}{M_s} \left[\vec{\Omega}\times\frac{\delta E}{\delta\vec{\Omega}}\right] + \eta_\ts{g}\left[\vec{\Omega}\times \partial_t\vec{\Omega}\right],
\end{equation}
where $\vec{\Omega}$ is unit magnetization vector, $\gamma_0$ is gyromagnetic ratio, and $\eta_\ts{g}$ is the Gilbert damping parameter. For the description of the skyrmion dynamics, we use a collective variable approach based on the traveling-wave Ansatz (TWA) formulated for the curvilinear coordinates: $\Omega_\alpha = \Omega_\alpha\left[x_1-X_1(t),x_2-X_1(t)\right]$, $\Omega_n =\Omega_n\left[x_1-X_1(t),x_2-X_1(t)\right]$,  where $\Omega_\alpha$ and $\Omega_n$ are the tangential and normal components of $\vec{\Omega}$, respectively. Note that $\vec{\Omega}\ne\vec{\Omega}\left[x_1-X_1(t),x_2-X_1(t)\right]$, in contrast to the case of a planar film. This is because the TWA is formulated for the magnetization components $\Omega_\alpha$, $\Omega_n$ defined in the coordinate-dependent basis. Substitution of the TWA into the Eq.~\eqref{eq:LLG} with the subsequent integration over the space domain results in the Thiele equation for collective coordinates \cite{Thiele73,Korniienko20}, which has the following normalized form
\begin{equation}\label{eq:Thiele-tensor-dmnls}	\left(N_{\text{top}}\varepsilon_{\alpha\beta}-\eta\mathcal{D}_{\alpha\beta}\right)\dot{\mathcal{X}}_\beta = \frac{\partial \mathcal{E}}{\partial \mathcal{X}_\alpha},
\end{equation}
for details see Appendix~\ref{app:dyn-fm}. Here we use the advantage of the orthonormalized basis and do not distinguish between the contra- and covariant indices. Units for energy and distance are the same as for Eq.~\eqref{eq:Ecrv}. The overdot means the derivative with respect to the dimensionless time $\tau=\omega_{\textsc{fm}}t$ with $\omega_{\textsc{fm}}=2K\gamma_0/M_s$. 
%Although Eq.~\eqref{eq:Thiele-tensor-dmnls} is valid for an arbitrary topological charge $N_{\text{top}}$ of the skyrmion, the results of the previous sections are obtained for a specific case $N_{\text{top}}=-1$. 
Damping in Eq.~\eqref{eq:Thiele-tensor-dmnls} is described by the re-scaled constant $\eta=\mathcal{C}_0\eta_\ts{g}$ and the damping tensor whose components are different for different skyrmion types:
\begin{equation}\label{eq:D-nb}
	\left[\mathcal{D}_{\alpha\beta}^{\textsc{n}}\right]\approx\begin{bmatrix}
		1-\delta^\ts{n} & 0 \\
		0 & 1+\delta^\ts{n}
	\end{bmatrix},\quad	\left[\mathcal{D}_{\alpha\beta}^{\textsc{b}}\right]\approx\begin{bmatrix}
		1 & \delta^\ts{b} \\
		\delta^\ts{b} & 1
	\end{bmatrix}.
\end{equation}
Here the linear in curvature corrections $\delta^{\ts{n},\ts{b}}=\frac{\mathcal{C}_2}{2\mathcal{C}_0}[g^{\ts{n},\ts{b}}(\alpha_g-1)-f^{\ts{n},\ts{b}}(\alpha_f-1)]$ appear due to the elliptical deformation of the skyrmion shape, and terms of the order $o(\varkappa)$ are neglected. Note that the corrections $\delta^{\ts{n},\ts{b}}$  are coordinate dependent, because of curvature dependence of the deformation amplitudes $f^{\ts{n},\ts{b}}$ and $g^{\ts{n},\ts{b}}$.

\begin{figure}
	\includegraphics[width=\columnwidth]{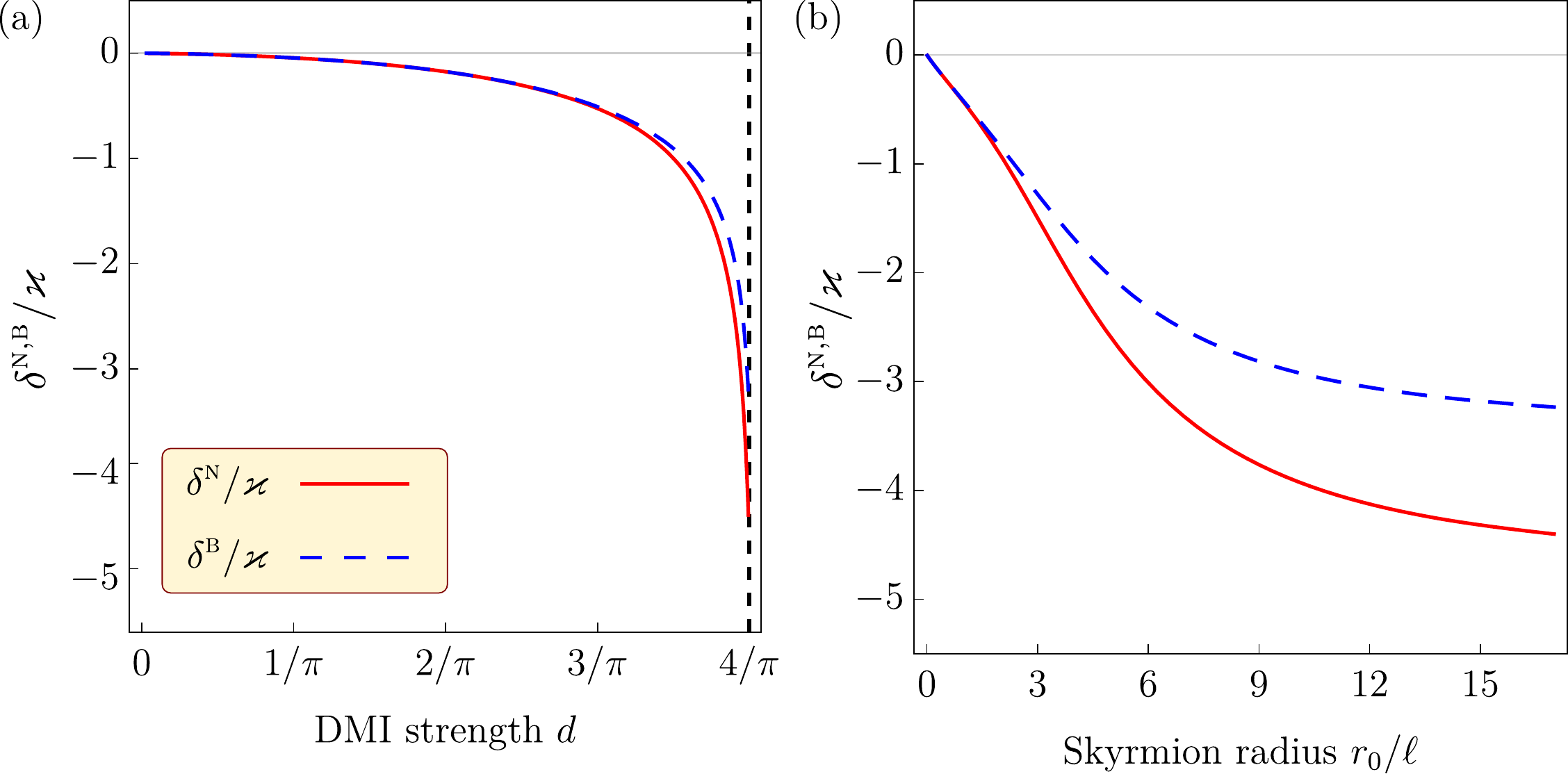}
	\caption{Correction $\delta^{\ts{n},\ts{b}}$ for the damping tensors \eqref{eq:D-nb} as functions of (a) the DMI constant and (b) the skyrmion radius.}\label{fig:D}
\end{figure}

Since the skyrmion effective energy depends on the single coordinate $\mathcal{E}=\mathcal{E}(\mathcal{X}_1)$, one can conclude from  \eqref{eq:Thiele-tensor-dmnls} that the skyrmion always moves along a trajectory $\x_1 - \x_1(0)\approx-\frac{\eta}{N_{\text{top}}}[\x_2 - \x_2(0)]$ which is linear in the curvilinear coordinates.
%\begin{equation}\label{eq:traj-fm}
%	\x^1 - \x^1(0)\approx-\frac{\eta_\ts{fm}}{N_{\text{top}}}\left[\x^2 - \x^2(0)\right].
%\end{equation}
Here and in what follows we assume that the damping is small and we neglect terms proportional to $\eta^2$ and $\eta\delta^{\ts{n},\ts{b}}$. Example of a skyrmion trajectory obtained as a numerical solution of Eqs.~\eqref{eq:Thiele-tensor-dmnls} for the case of a Bloch skyrmion is shown in Fig.~\ref{fig:sine_geometry}a. 
The skyrmion velocity
\begin{equation}\label{eq:fm_sk_vel}
	\vec{V}_\ts{fm}\approx \frac{\mathcal{E}'(\x_1) }{N_{\text{top}}^2} \left[ -\eta\vec{e}_1 +N_{\text{top}}\vec{e}_2\right],
\end{equation}
with $\mathcal{E}'(\x_1)=\partial \mathcal{E}/\partial \x_1$ is determined by the skyrmion position only. For the limit case of vanishing damping ($\eta_\ts{fm}\to 0$) skyrmion moves along the cylinder generatrix with velocity $\vec{V}_\ts{fm} = \hat{\vec{z}}\mathcal{E}'(\x_1)/N_{\text{top}}$. The generatrices $\{\x_1=\x^0_1,\,\x_2\in\mathbb{R}\}$ where $\x^0_1:\,\mathcal{E}'(\x^0_1)=0$ are stationary lines. Linearization of the equations of motion \eqref{eq:Thiele-tensor-dmnls} in vicinity of the stationary lines results in solution $	\tilde{\x}_1(\tau)\approx \tilde{\x}_1(0)\exp[-\eta N_{\text{top}}^{-2}\mathcal{E}''(\x_1^0)\tau]$,
%\begin{equation}\label{eq:x1-tilde}
%	\tilde{\x}^1(\tau)\approx \tilde{\x}^1(0)\exp\left[-\frac{\eta_\ts{fm}}{N_{\text{top}}^2}\mathcal{E}''(\x^1_0)\tau\right],
%\end{equation}
where $\tilde{\x}_1=\x_1-\x^0_1$ is small displecement from the stationary line. Thus, one can conclude that the statinary line is stable if $\mathcal{E}''(\x^0_1)>0$, and unstable if $\mathcal{E}''(\x^0_1)<0$. Taking into account the form of the energy \eqref{eq:Ecrv} for different types of skyrmions, one can reduce the obtained stability condition to \eqref{eq:stbl-eqlbr-N} and \eqref{eq:stbl-eqlbr-B} for N{\'e}el and Bloch skyrmions, respectively.

\subsection{FM skyrmion in the cylindrical Euler spiral.}

\begin{figure*}
	\includegraphics[width=\textwidth]{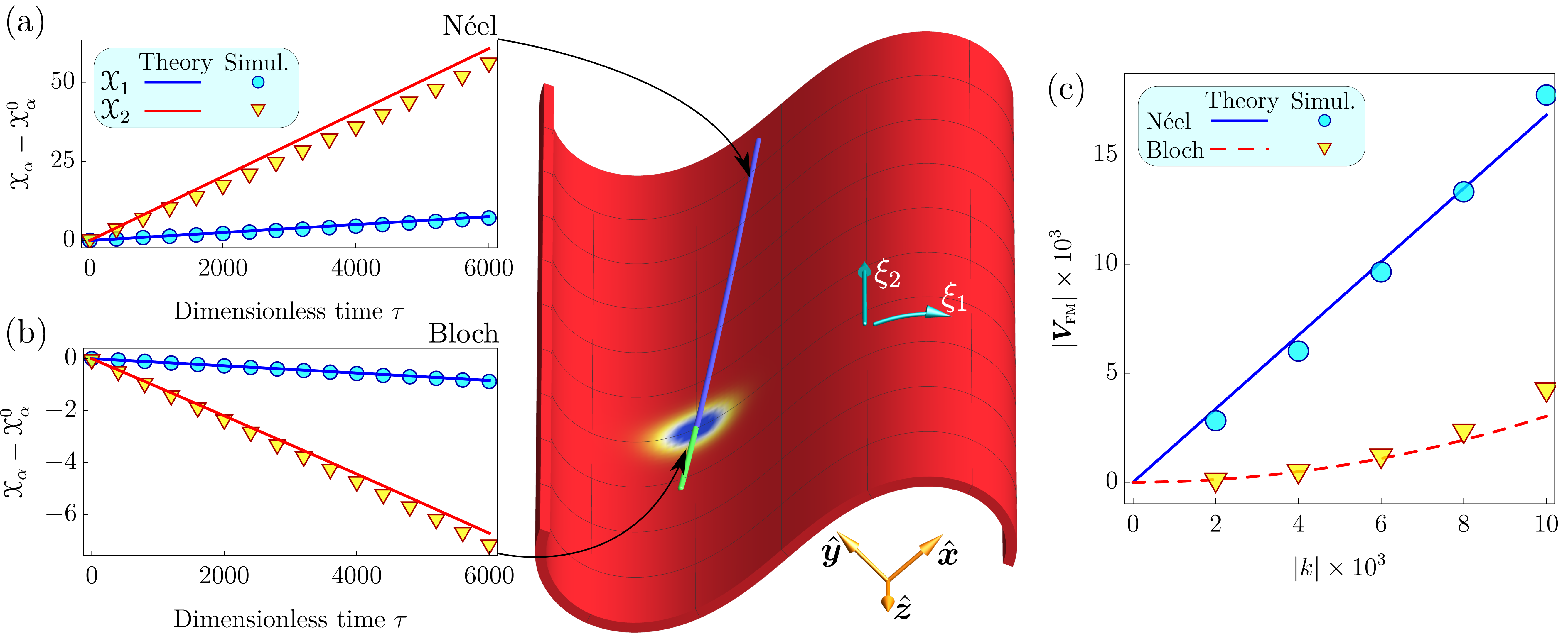}
	\caption{\label{fig:coord_FM}%
		(Color online) Curvature-induced drift of FM skyrmions along cylindrical Euler spiral. Time dependencies of coordinates of N{\'e}el (panel a) and Bloch (panel b) FM skyrmions obtained by means of micromagnetic simulations (markers) are compared with analytical predictions of the collective variables model (solid lines), namely coordinates of N{\'e}el and Bloch skyrmions are obtained by means of Eq.~\eqref{eq:fm_neel_cornu_tr} and Eq.~\eqref{eq:fm_bloch_cornu_tr}, respectively. Skyrmion trajectories on central panel are extracted from simulations, see Appendix~\ref{app:simuls} for details. In simulations, velocities of skyrmions are determined as explained in Appendix~\ref{app:simuls}. For simulations we use $d=1$, $k=6\times 10^{-3}$,  $\eta_\ts{g} = 0.1$, initial skyrmions positions are $\x_1(0) = -15$ and $\x_2(0) = 0$. Dynamics of both skyrmions is shown in supplemental movies \cite{Note2}. Panel (c) demonstrates dependence of skyrmion velocity on the curvature gradient. Markers represent the velocity averaged during the simulation time $\tau_{\text{sim}}=6\times10^3$. 
		 Solid and dashed lines in (c) are analytical estimations of velocities of N{\'e}el $|\vec{V}_\ts{fm}^\ts{n}|\approx k|\mathfrak{C}_1^{\ts{n}}|$ and Bloch $|\vec{V}_\ts{fm}^\ts{b}| \approx k^2 |\mathfrak{C}^\ts{b}_2\x_1(0)|$ skyrmions, respectively. Note that the approximation of the constant velocity for the Bloch skyrmion is valid since $\tau_s\ll\tau^*\sim10^4-10^5$.}
\end{figure*} 

As an example we consider a cylindrical surface with a shape of Euler spiral (also known as a Cornu spiral or clothoid)~\cite{Lawrence14}, it is determined by the directrix
\begin{equation}\label{eq:Euler_spiral}
\vec{\gamma}(\xi_1) = \sqrt{\frac{2}{k}}\left[ \hat{\vec{x}}\,\mathrm{C}\left(\sqrt{\frac{k}{2}}\xi_1\right)-\hat{\vec{y}}\,\mathrm{S}\left(\sqrt{\frac{k}{2}}\xi_1\right)\right],
\end{equation}
where $\mathrm{C}(x) = \int_{0}^{x}\cos (y^2)\mathrm{d}y$ and $\mathrm{S}(x) = \int_{0}^{x}\sin (y^2)\mathrm{d}y$ are the Fresnel integrals, see Fig.~\ref{fig:coord_FM}. The directrix curvature $\varkappa=k\xi_1$ has constant gradient, so the cylindrical Euler spiral is the simplest surface for analysis of the curvature induced skyrmion drift. 

Energy of the \emph{N{\'e}el} skyrmion is approximated as $\mathcal{E}(\x_1)\approx-|\mathfrak{C}_1^{\ts{n}}|k\x_1+\text{const}$, where we take into account that $\mathfrak{C}_1^{\ts{n}}<0$ assuming $D>0$. In this case the equations of motion \eqref{eq:Thiele-tensor-dmnls} result into the skyrmion trajectory $\vec{\x}(\tau)=\vec{\varsigma}(\x_1(\tau),\x_2(\tau))$ with $	\x_1(\tau)\approx\x_1(0)+|\mathfrak{C}_1^{\ts{n}}|k\eta\tau$ and $\x_2(\tau)\approx\x_2(0)+|\mathfrak{C}_1^{\ts{n}}|k\tau$. Here we substitute $N_{\text{top}}=-1$, because the quantity $\mathfrak{C}^{\textsc{n}}_1$ obtained in Section~\ref{sec:stat-skyr} and shown in Fig.~\ref{fig:Cn-Cb} is valid only for this case.
%\begin{equation}\label{eq:fm_neel_cornu_tr}
%	\x^1(\tau)\approx\x^1(0)-\frac{\mathfrak{C}_1^{\ts{n}}k\eta_{\ts{fm}}}{N_{\text{top}}^2}\tau,\qquad \x^2(\tau)\approx\x^2(0)+\frac{\mathfrak{C}_1^{\ts{n}}k}{N_{\text{top}}}\tau.
%\end{equation}
The predicted trajectory is in good agreement with micromagnetic simulations, Fig.~\ref{fig:coord_FM}a. Note that due to small damping, N{\'e}el skyrmion moves mainly along generatrix.
Absolute value of skyrmion velocity is approximately a constant $|\vec{V}_\ts{fm}^\ts{n}|\approx k|\mathfrak{C}_1^{\ts{n}}|$, which is linear in the curvature gradient, see Fig.~\ref{fig:coord_FM}c. Here the terms $o(\eta_\ts{fm})$ are neglected.
Since $\mathcal{E}'\ne0$, there are no stationary lines for N{\'e}el skyrmios on the cylindrical Euler spiral.

Energy of a \emph{Bloch} skyrmion on the cylindrical Euler spiral is approximated as $\mathcal{E}(\x_1)\approx-\frac12|\mathfrak{C}_2^{\ts{b}}|k^2\x_1^2+\text{const}$. For this case, the equations of motion \eqref{eq:Thiele-tensor-dmnls} result into the skyrmion trajectory $\vec{\x}(\tau)$ with coordinates \eqref{eq:fm_bloch_cornu_tr}.
%\begin{equation}\label{eq:fm_bloch_cornu_tr}
%	\x^1 \approx \x^1(0)\text{exp}\left(\dfrac{\eta_\ts{fm}|\mathfrak{C}_2^\ts{b}|\, k^2}{N_{\text{top}}^2}\tau\right),\qquad 
%	\x^2 \approx \x^2(0)+\frac{\x^1(0)N_{\text{top}}}{\eta_\ts{fm}}\left[1-\exp\left(\dfrac{\eta_\ts{fm}|\mathfrak{C}_2^\ts{b}|\, k^2}{N_{\text{top}}^2}\tau\right)\right].
%\end{equation}
In contrast to N{\'e}el skyrmion, velocity \eqref{eq:fm_bloch_cornu_vel} of a Bloch skyrmion is not a constant.  
%\begin{equation}\label{eq:fm_bloch_cornu_vel}
%	|\vec{V}_\ts{fm}^\ts{b}| \approx \frac{|\x^1(0)\mathfrak{C}^\ts{b}_2|\, k^2}{|N_{\text{top}}|}\exp\left(\dfrac{\eta_\ts{fm}|\mathfrak{C}_2^\ts{b}|\, k^2}{N_{\text{top}}^2}\tau\right).
%\end{equation}
%One can see, that velocity of Bloch skyrmion depends on the initial position. And for the skyrmion position with $\x_1(0) = 0$, or $\h[\x_1(0)]=0$, the skyrmion is immobile. 
However, for initial moments of time $\tau\ll\tau^*$, where  $\tau^*=1/(\eta|\mathfrak{C}^\ts{b}_2|k^2)$, one can approximate $|\vec{V}_\ts{fm}^\ts{b}| \approx |\mathfrak{C}^\ts{b}_2\x_1(0)|\, k^2$. In contrast to N{\'e}el skyrmion, velocity of Bloch skyrmion is quadratic in the curvature gradients and, for this reason, it is much smaller see Fig.~\ref{fig:coord_FM}(c). Note also different scales of the vertical axes in panels (a) and (b). 

Dynamics of a Bloch skyrmion essentially depends on its initial position $\x_1(0)$. In Fig.~\ref{fig:coord_FM} we choose $\x_1(0)<0$, for this reason, Bloch skyrmion moves in the direction opposite to the N{\'e}el skyrmion. In contrast, direction of motion of the N{\'e}el skyrmion is independent on its initial position.

For a Bloch skyrmion there is single unstable stationary line $\x_1^0=0$ on the surface \eqref{eq:Euler_spiral}. One can see from \eqref{eq:fm_bloch_cornu_vel} that $V_\ts{fm}^\ts{b}=0$ if $\x_1(0)=0$.

%The direction of motion for the FM N{\'e}el skyrmion is defined by the sign of product of the curvature gradient and DMI constant: $k \mathfrak{C}^\ts{n}_1(d)$, while for the FM Bloch skyrmion the direction is defined only by the sign of the initial position $\x^1(0)$. The corresponding time evolution of the FM skyrmion positions $\x^\alpha$ for N{\'e}el and Bloch are presented in Figs.~\ref{fig:coord_FM}(a) and~\ref{fig:coord_FM}(b), respectively.

\section{Curvature induced dynamics of AFM skyrmions} \label{sec:dynamics-AFM}

Within the exchange approximation  a two-sublattices fully compensated AFM can be described by the  N{\'e}el order parameter  $\vec{\Omega}$ whose dynamics is determined by the equation \cite{Baryakhtar79,Ivanov95e,Turov01en}
\begin{equation}\label{eq:n}
	\left[\frac{2M_s}{\gamma_0^2B_{\ts{x}}}\partial_t^2{\vec{\Omega}}+\frac{\delta E}{\delta\vec{\Omega}}+\eta_\ts{g}\frac{2M_s}{\gamma_0}\partial_t{\vec{\Omega}}\right]\times\vec{\Omega}=0.
\end{equation}
Here Hamiltonian $E$ is defined in \eqref{eq:model}, and it is assumed that the exchange field $B_\ts{x}$ \cite{Note1} much exceeds all the other effective fields in the system, e.g. $B_\ts{x}\gg B_\ts{a}$ with $B_\ts{a}=K/M_s$ being the anisotropy field. 

Applying TWA to \eqref{eq:n} we proceed to the equation of motion for the collective coordinates
\begin{equation}\label{eq:eqs-afm}
	\mathcal{M}_{\alpha\beta}\left(\ddot{\mathcal{X}}_\beta+\bar{\eta}\dot{\mathcal{X}}_\beta\right)=-\frac{\partial\mathcal{E}}{\partial\x_\alpha},\qquad \mathcal{M}_{\alpha\beta}=\mathcal{C}_0\mathcal{D}_{\alpha\beta}.
\end{equation}
For details see Appendix~\ref{app:dyn-afm}. Here overdot denotes derivative with respect to dimensionless time $\bar{\tau}=t\omega_{\textsc{afm}}$, where $\omega_{\ts{afm}}=\gamma_0\sqrt{B_\ts{x}B_\ts{a}}$ is frequency of the uniform antiferromagnetic resonance. The rescaled damping coefficient is $\bar\eta=\sqrt{B_\ts{x}/B_\ts{a}}\eta_{\ts{g}}$. In contrast to the FM case, the role of tensor $\mathcal{D}_{\alpha\beta}$ is significant in Eq.~\eqref{eq:eqs-afm} even in the small-damping limit, since it determines the tensor of skyrmion mass $\mathcal{M}_{\alpha\beta}$. Since tensor $\mathcal{D}_{\alpha\beta}$ has different structure for the N{\'e}el and Bloch skyrmions, these cases must be considered independently.

{\bf N{\'e}el skyrmion.} Utilizing the fact that the energy $\mathcal{E}$ in \eqref{eq:eqs-afm} depends on single coordinate $\x_1$, we present the equations of motion in form
\begin{subequations}\label{eq:eqs-afm-N}
	\begin{align}
		\label{eq:X1-N}&\ddot{\x}_1+\bar\eta\dot{\x}_1+\frac{1}{\mathcal{C}_0}\mathcal{E}'(\x_1)\approx0,\\
		\label{eq:X2-N}&\ddot{\x}_2+\bar\eta\dot{\x}_2=0,
	\end{align}
\end{subequations}
where Eq.~\eqref{eq:X1-N} is written in the leading approximation in curvature. Eq.~\eqref{eq:X2-N} has general solution $\x_2(\bar\tau)=\x_2(0)+V_2(0)(1-e^{-\bar\eta\bar\tau})/\bar\eta$,
%\begin{equation}\label{eq:X2-N-gen}
%	\x^2(\tau)=\x^2(0)+V^2(0)\frac{1-e^{-\eta_{\ts{afm}}\tau}}{\eta_{\ts{afm}}},
%\end{equation}
where $V_\alpha=\dot{\x}_\alpha$. So, for zero initial velocity one has $\x_2=\text{const}$ and skyrmion moves only in direction $\vec{e}_1$, i.e. along directrix. Equation \eqref{eq:X1-N} results in stationary lines $\x_1=\x^0_1$ determined by equation $\mathcal{E}'(\x_1^0)=0$. The stationary line is stable (unstable) if $\mathcal{E}''(\x_1^0)>0$ ($\mathcal{E}''(\x_1^0)<0$). For the energy \eqref{eq:ser-N}, the latter stability condition is equivalent to \eqref{eq:stbl-eqlbr-N} in the leading in curvature approxiamtion. 

\begin{figure}
	\includegraphics[width=0.8\columnwidth]{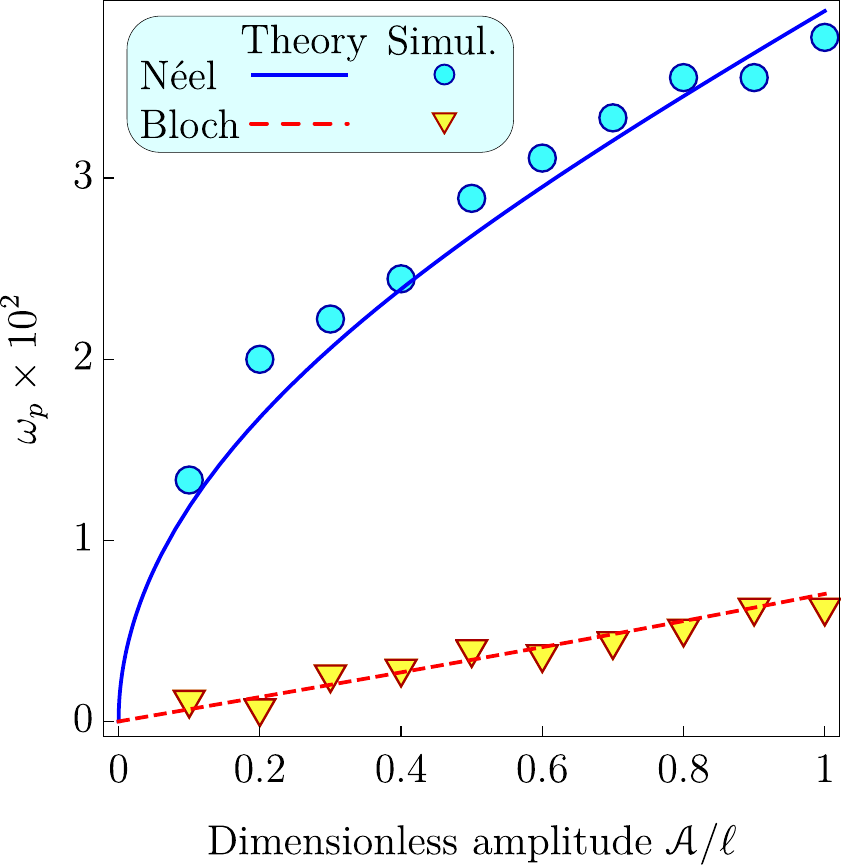}
	\caption{\label{fig:oscillations}%
		(Color online) Eigenfrequency of oscillations of the AFM skyrmion in vicinity of the stable equilibrium for the case of sinusoidal cylinder with generatrix \eqref{eq:sin}. Solid and dashed lines correspond to the eigenfrequencies of N{\'e}el and Bloch skyrmions plotted for predictions \eqref{eq:w-Neel} and \eqref{eq:w-Bloch}, respectively. Symbols correspond to the data obtained by means of numerical simulations for sinusoidal film with $\mathcal{T} = 35\ell$, $d=1$, and $\eta_\ts{g} = 10^{-4}$.}
\end{figure}

In vicinity of a stable stationary line skyrmion demonstrates the oscillatory dynamics $\x_1(\bar\tau)\approx \x_1^0+ a\cos(\omega_p^\ts{n}\bar\tau+\psi_0)e^{-\bar\eta\bar\tau/2}$ and $\x_2(\bar\tau)=\text{const}$ with frequency $\omega_p^\ts{n}=\sqrt{\mathcal{C}_0^{-1}\mathcal{E}''(\x^0_1)}$.  In the leading in curvature approximation, one obtains from \eqref{eq:ser-N}
\begin{equation}\label{eq:w-Neel}
	\omega_p^\ts{n}=\sqrt{\frac{|\mathfrak{C}_1^\ts{n}|}{\mathcal{C}_0}|\varkappa''(\x^0_1)|}.
\end{equation}
Note that frequency is measured in units of $\omega_{\ts{afm}}$.
For the sinusoidal surface with generatrix \eqref{eq:sin} the stable stationary lines are $x_0=(1/4+n)2\pi/q$ with $\omega_p^\ts{n}=\sqrt{|\mathfrak{C}_1^\ts{n}|\mathcal{C}_0^{-1}\mathcal{A}\ell^3q^4(1+3\mathcal{A}^2q^2)}$. The obtained frequency is in good agreement with micromagnetic simulations, see Fig.~\ref{fig:oscillations} and the supplemental movie \cite{Note2}. 

%======================================================================================================================
%======================================================================================================================

{\bf Bloch skyrmion.} Assuming that the damping is small, we neglect terms of the order $\bar\eta^2$, $\bar\eta\delta^\ts{b}$ and higher. For this case, we approximate Eqs.~\eqref{eq:eqs-afm} as follows
\begin{subequations}\label{eq:eqs-afm-rs}
	\begin{align}
		\label{eq:X1}&\ddot{\x}_1+\bar\eta\dot{\x}_1+\frac{1}{\mathcal{C}_0}\mathcal{E}'(\x_1)\approx0,\\
		\label{eq:X2}&\ddot{\x}_2+\bar\eta\dot{\x}_2-\frac{\delta^\ts{b}(\x_1)}{\mathcal{C}_0}\mathcal{E}'(\x_1)\approx0.
	\end{align}
\end{subequations}
In contrast to the equations of motion \eqref{eq:eqs-afm-N} for the N{\'e}el skyrmion, Eq.~\eqref{eq:X2} for the coordinate $\x_2$ obtains a driving determined by the solution for $\x_1(\bar\tau)$. This means that the Bloch skyrmion always experiences a drift in $\vec{e}_2$ direction, in contrast to the N{\'e}el skyrmion.

For the case of Bloch skyrmion in the leading in curvature  approximation we have $\mathcal{E}=-\frac12|\mathfrak{C}_2^\ts{b}|\varkappa^2(\x_1)$ and the stable equilibrium lines are determined by \eqref{eq:stbl-eqlbr-B}. In vicinity of a stable stationary line Bloch skyrmion demonstrates oscillations  $\x_1(\bar\tau)\approx \x_1^0+ a\cos(\omega_p^\ts{b}\bar\tau+\psi_0)e^{-\bar\eta\bar\tau/2}$ and $\x_2(\bar\tau)=\delta^\ts{b}(\x^0_1)a\cos(\omega_p^\ts{b}\bar\tau+\psi_0)e^{-\bar\eta\bar\tau}$ with frequency
\begin{equation}\label{eq:w-Bloch}
	\omega_p^\ts{b}=\sqrt{\frac{|\mathfrak{C}_2^\ts{b}|}{\mathcal{C}_0}|\varkappa'(\x^0_1)^2+\varkappa''(\x^0_1)\varkappa(\x^0_1)|}.
\end{equation}
In the local basis $\{\vec{e}_1,\vec{e}_2\}$, this solution corresponds to the oscillations along a straight line which makes angle $\alpha=\arctan\delta^\ts{b}(\x^0_1)$ with the direction $\vec{e}_1$. For the sinusoidal surface with generatrix \eqref{eq:sin} the stable stationary lines are $x_0=(1/4+n/2)2\pi/q$ with $\omega_p^\ts{b}=\sqrt{|\mathfrak{C}_2^\ts{b}|\mathcal{C}_0^{-1}\mathcal{A}^2\ell^4q^6(1+3\mathcal{A}^2q^2)}$. The obtained frequency is in good agreement with micromagnetic simulations, see Fig.~\ref{fig:oscillations} and the supplemental movie \cite{Note2}.   

Note the drastically different behavior in the small amplitude limit: $\omega_p^\ts{n}\propto\sqrt{\mathcal{A}}$ and $\omega_p^\ts{b}\propto\mathcal{A}$.
\subsection{AFM skyrmion in the cylindrical Euler spiral.}
\begin{figure*}
	\includegraphics[width=\textwidth]{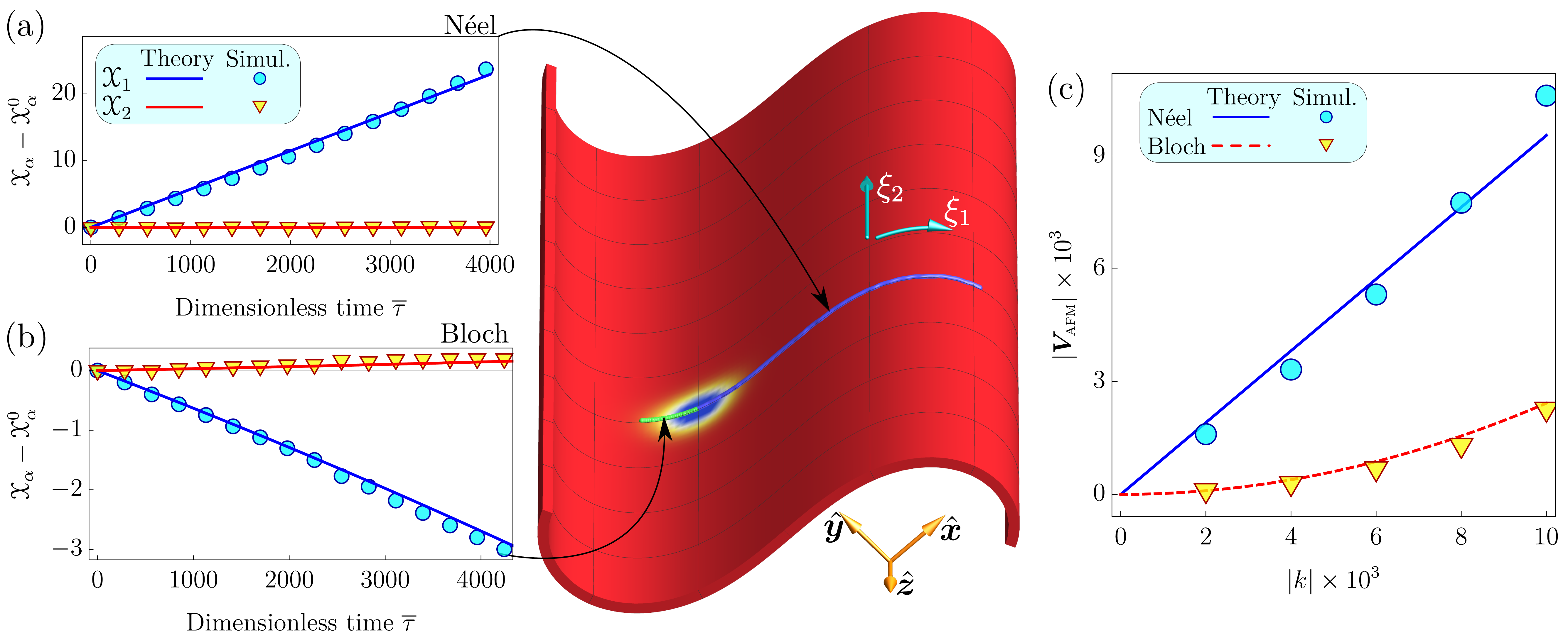}
	\caption{\label{fig:coord_AFM}%
				(Color online) Curvature-induced drift of AFM skyrmions along cylindrical Euler spiral. Time dependencies of coordinates of N{\'e}el (panel a) and Bloch (panel b) FM skyrmions obtained by means of micromagnetic simulations (markers) are compared with analytical predictions of the collective variables model (solid lines), namely coordinates of N{\'e}el and Bloch skyrmions are obtained by means of Eq.~\eqref{eq:X1-sol} and Eqs.~(\ref{eq:X1-ksmall}, \ref{eq:X2-ksmall}), respectively. Skyrmion trajectories on central panel are extracted from simulations, see Appendix~\ref{app:simuls} for details.  For simulations we use $d=1$, $k=6\times 10^{-3}$,  $\eta_\ts{g} = 0.1$, initial skyrmions positions are $\x_1(0) = -15$ and $\x_2(0) = 0$, and initial velocity vanishes.  Dynamics of both skyrmions is shown in supplemental movies \cite{Note2}. Panel (c) demonstrates the skyrmion velocities in the long time limit $\bar\tau\gg1/\bar\eta$. Solid and dashed lines in (c) are absolute values of velocities of N{\'e}el $|\vec{V}_\ts{afm}^\ts{n}|\approx k|\mathfrak{C}_1^{\ts{n}}|/(\bar{\eta}\mathcal{C}_0)$ and Bloch $|\vec{V}_\ts{afm}^\ts{b}| \approx k^2 |\mathfrak{C}^\ts{b}_2\x_1(0)|/(\mathcal{C}_0\bar{\eta})$ skyrmions, respectively, in the limit of small curvature gradients $k\ll\bar\eta$. Markers shows the velocities extracted from the simulations and averaged during the simulation time.}
\end{figure*} 

Trajectory of a \emph{N{\'e}el} skyrmion can be found as a solution of equations of motion \eqref{eq:eqs-afm-N} for the energy \eqref{eq:ser-N} with the form of curvature $\varkappa=k\x_1$ taken into account. If the curvature gradients are small $k\ll\bar{\eta}$, then the N{\'e}el skyrmion demonstrates a uniformly accelerated motion with the acceleration $\vec{a}^{\textsc{n}}=-\mathfrak{C}^\ts{n}_1k/\mathcal{C}_0\vec{e}_1$ for the initial moments of time $\bar\tau\ll1/\bar\eta$, and it reaches the regime with constant velocity $\vec{V}_\ts{afm}^\ts{n}\approx\vec{a}^{\textsc{n}}/\bar\eta$ in the opposite limit of long times $\bar\tau\gg1/\bar\eta$, see Appendix~\ref{app:dyn-afm} for details. Here it was assumed that  the initial skyrmion velocity vanishes.

In the same manner, one obtains the trajectory for \emph{Bloch} skyrmion solving equations of motion \eqref{eq:eqs-afm-rs} for the energy \eqref{eq:ser-B}. Similarly to the N{\'e}el skyrmion, in the limit of small curvature gradients, Bloch skyrmion demonstrates a uniformly accelerated motion with the acceleration $\vec{a}^{\textsc{b}}=\x_1(0)k^2|\mathfrak{C}^\ts{b}_1|\mathcal{C}_0^{-1}[\vec{e}_1+\x_1(0)k\Delta^\ts{b}\vec{e}_2]$  for the initial moments of time, and it reaches the regime with constant velocity $\vec{V}_\ts{afm}^\ts{b}\approx\vec{a}^{\textsc{b}}/\bar\eta$, in the opposite limit of long times $\bar\tau\gg1/\bar\eta$, see Appendix~\ref{app:dyn-afm} for details. Here $\Delta^\ts{b}=\delta^\ts{b}/\varkappa$, see Fig.~\ref{fig:D} and zero initial velocity was assumed.

Thus, in the long time limit, both skyrmions reach the regime of constant velocity which is linear and quadratic in $k$ (curvature gradient) for N{\'e}el and Bloch skyrmions, respectively. The Bloch skyrmion moves much slower, in agreement with our simulations, see Fig.~\ref{fig:coord_AFM}. Both skyrmions mainly move along directrix (in direction $\vec{e}_2$), however, Bloch skyrmion obtains small velocity component along generatrix (direction $\vec{e}_1$), see Fig.~\ref{fig:coord_AFM}b. The latter effect is attributed to the elliptical skyrmion deformation, which results in a weak nondiagonality of the mass tensor.

\section{Conclusions}

We have built a theory of curvature-induced drift of magnetic (FM and AFM) skyrmions along curvilinear surfaces of cylindrical geometry. N{\'e}el skyrmions experience a much stronger influence of the curvature gradients and have larger drift velocities compared to Bloch skyrmions. Both types of skyrmions experience curvature-induced deformations of their profiles. This deformation has different impact on the skyrmion dynamics: significantly influences dynamics of the slow Bloch skyrmions, while it can be neglected for the description of dynamics of fast N{\'e}el skyrmions. We have also developed the theory of skyrmion deformations on cylindrical surfaces. Two types of deformations can be distinguished: radially-symmetric and elliptical ones.

The predicted redistribution of density of skyrmion gas along the sinusoidally deformed film, see Fig.~\ref{fig:sine_geometry}, does not require strong curvature gradients and, therefore, it can be used for an experimental verification of the curvature induced skyrmion drift. 

\section{Acknowledgments}
In part, this work was supported by the Program of Fundamental Research of the Department of Physics and Astronomy of the National Academy of Sciences of Ukraine (Project No. 0117U000240) and the National Research Foundation of Ukraine (Project No. 2020.02/0051). We thank Ulrike Nitzsche for technical support.

\appendix

\section{Energy of a rigid skyrmion in curved film}\label{app:energy}
In terms of the angular parameterization \eqref{eq:ang-par}, the exchange energy density is \cite{Gaididei14} $\mathscr{E}_\textsc{x}=\left[\vec{\nabla}\theta-\vec{\Gamma}\right]^2+[\sin\theta\,\vec{\nabla}\phi-\cos\theta\partial_\phi\,\vec{\Gamma}]^2$,
%	\begin{equation}\label{eq:exchnage_angle}
%	\mathscr{E}_\textsc{x}=\left[\vec{\nabla}\theta-\vec{\Gamma}\right]^2+\left[\sin\theta\,\vec{\nabla}\phi-\cos\theta\partial_\phi\,\vec{\Gamma}\right]^2,
%\end{equation}
where $\vec{\Gamma}=\vec{e}_\alpha h_{\alpha\beta}\varepsilon_\beta$ with $h_{\alpha\beta}$ being the Weingarten map. For the case under consideration $h_{\alpha\beta}=\text{diag}(-\kappa,0)$ and $\vec{\Gamma}=-\vec{e}_1\kappa\cos\phi$. The DMI energy densities are $	\mathscr{E}_\textsc{d}^\textsc{n}=2\left(\vec{\nabla}\theta\cdot\vec{\varepsilon}\right)\sin^2\theta+\kappa\cos^2\theta$ and $	\mathscr{E}_\textsc{d}^\textsc{b} = \sin^2\theta\,\left[\left(2\vec{\nabla}\theta-\vec{\Gamma}\right)\times\vec{\varepsilon}\right]\cdot\vec{n}$ for the interfacial (N{\'e}el) \cite{Kravchuk18a} and isotropic (Bloch) \cite{Yershov19a} DMI, respectively. The anisotropy energy density is trivial $\mathscr{E}_\textsc{a}=1-\Omega_n^2=\sin^2\theta$.

%\subsection{Rigid skyrmion}

For Ansatz \eqref{eq:Ansatz-rigid} the exchange energy density obtains the form
\begin{equation}\label{eq:Eex-den-Neel}
	\begin{split}
	\mathscr{E}_\textsc{x}^\textsc{n}=&\,\theta_0'^2+\frac{\sin^2\theta_0}{r^2}+\kappa^2\left[1-\sin^2\chi\sin^2\theta_0\right]\\
	&+\sigma\kappa\left[2\theta_0'\cos^2\chi+\frac{\sin2\theta_0}{r}\sin^2\chi\right]
	\end{split}
\end{equation}
and
\begin{equation}\label{eq:Eex-den-Bloch}
	\begin{split}
	\mathscr{E}_\textsc{x}^\textsc{b}=&\theta_0'^2+\frac{\sin^2\theta_0}{r^2}+\kappa^2\left[1-\cos^2\chi\sin^2\theta_0\right]\\
	&+\sigma\kappa\sin2\chi\left[-\theta_0'+\frac{\sin\theta_0\cos\theta_0}{r}\right]
	\end{split}
\end{equation}
for N{\'e}el and Bloch skyrmions, respectively. Integration over the film area results in the following total exchange energies $E_\textsc{x}^{\textsc{n},\textsc{b}}=AL\iint\mathscr{E}_\textsc{x}^{\textsc{n},\textsc{b}}\mathrm{d}x_1\mathrm{d}x_2$:
\begin{equation}\label{eq:Eex-Neel}
	\begin{split}
	E_\textsc{x}^{\textsc{n}}=8\pi AL\biggl\{C_0&+\sigma\kappa(X_1)C_1-\frac12\kappa^2(X_1)C_2\\
	&+\mathcal{O}(\kappa''r_0^3)+F[\kappa]\biggr\}
	\end{split}
\end{equation}
and 
\begin{equation}\label{eq:Eex-Bloch}
	E_\textsc{x}^{\textsc{b}}=8\pi AL\left\{C_0-\frac12\kappa^2(X_1)C_2+\mathcal{O}(\kappa'^2r_0^4,\kappa\kappa''r_0^4)+F[\kappa]\right\}.
\end{equation}
In order to obtain \eqref{eq:Eex-Neel} and \eqref{eq:Eex-Bloch} we present the curvature in the form of series \eqref{eq:kappa-ser}
and utilize the fact that functions $\sin\theta_0(r)$ and $\theta_0'(r)$ are exponentially localized, and use the approximation \eqref{eq:limit}. Note that the area element $\dd\mathcal{S}=\dd x_1\dd x_2$ because of the Euclidean metric.
Functional $F[\kappa]=\frac14\iint\kappa^2\mathrm{d}x_1\mathrm{d}x_2$ is independent on the skyrmion position. Constants $C_n$ are as follows
\begin{equation}\label{eq:Cn}
	\begin{split}
	&C_0=\frac14\int\limits_{0}^\infty\left[\theta_0'(r)^2+\frac{\sin^2\theta_0(r)}{r^2}\right]r\dd r,\\ &C_1=\frac14\int\limits_{0}^\infty\left[\theta_0'(r)+\frac{\sin\theta_0(r)\cos\theta_0(r)}{r}\right]r\dd r,\\ &C_2=\frac14\int\limits_{0}^\infty\sin^2\theta_0(r)r\dd r.
	\end{split}
\end{equation}
Using the scaling transformation $\theta_0(r)\to\theta_0(r/r_0)$ one can roughly estimate $C_n\propto r_0^n$, where $r_0$ is skyrmion radius.

For Ansatz \eqref{eq:Ansatz-rigid} the DMI energy densities obtain form 
\begin{equation}\label{eq:Edmi-den}
	\begin{split}
	&\mathscr{E}_\textsc{d}^\textsc{n}=\sin^2\theta_0\left[2\sigma\theta_0'-\kappa\right]+\kappa,\\ &\mathscr{E}_\textsc{d}^\textsc{b}=\sin^2\theta_0\left[2\sigma\theta_0'-\kappa\sin\chi\cos\chi\right].
	\end{split}
\end{equation}
Under the same assumptions as above we obtain for the total DMI energies $E_\textsc{d}^{\textsc{n},\textsc{b}}=DL\iint\mathscr{E}_\textsc{d}^{\textsc{n},\textsc{b}}\mathrm{d}x_1\mathrm{d}x_2$:
\begin{equation}\label{eq:Edmi-tot-rigid}
	\begin{split}
	&E_\textsc{d}^{\textsc{n}}=8\pi DL\left\{\sigma C_1-\kappa(X_1)C_2+\mathcal{O}(\kappa''r_0^4)+\bar{F}[\kappa]\right\},\\
	&E_\textsc{d}^{\textsc{b}}=8\pi|D|LC_1
	\end{split}
\end{equation}
Here functional $\bar{F}[\kappa]=\frac14\iint\kappa\mathrm{d}x_1\mathrm{d}x_2$ is independent on the skyrmion position.

Skyrmions of both types have the same anisotropy energy $E_\textsc{a}=8\pi KL C_2$. 

Applying Derick's scaling transformations \cite{Leonov16} to \eqref{eq:model}, one obtains the following virial relation between different energy contributions $E_{\textsc{d}}+2E_{\textsc{a}}=0$ of the planar skyrmion. This results in the relation
\begin{equation}\label{eq:c1c2}
	|D|C_1=-2KC_2
\end{equation}
which enables us to exclude constant $C_1$ from consideration. Summing up the obtained energy contributions we present the total skyrmion energy in form of a sum of the planar and curvilinear parts, see Eqs.~\eqref{eq:Ecrv} and the discussion above. Note that for the dimensionless constants $\mathcal{C}_n=C_n/\ell^n$ the virial relation has the form $|d|\mathcal{C}_1=-2\mathcal{C}_2$.

\emph{Corrections of higher order in curvature.}
Condition \eqref{eq:limit} is violated in vicinity of special lines on the surface where $\kappa=0$. If a skyrmion crosses such a kind of line during its dynamics the higher order corrections in curvature must be taken into account. Performing the same procedure as described above we obtain for \eqref{eq:Ecrv}
\begin{subequations}\label{eq:E-high-order}
	\begin{align}
		\label{eq:E-high-orderN}&\mathcal{E}^{\textsc{n}}_{\text{crv}}\approx\mathfrak{C}^{\textsc{n}}_1\varkappa+\mathfrak{C}^{\textsc{n}}_2\frac{\varkappa^2}{2}+\frac{\mathcal{B}_1}{8}\sigma\varkappa''-\frac{\mathcal{B}_2}{8}\left[\varkappa'^2+\varkappa''(\varkappa+2d)\right],\\
		\label{eq:E-high-orderB}&\mathcal{E}^{\textsc{b}}_{\text{crv}}\approx\mathfrak{C}^{\textsc{b}}_2\frac{\varkappa^2}{2}-\frac38\mathcal{B}_2\left(\varkappa'^2+\varkappa\varkappa''\right)
	\end{align}
\end{subequations}
where coefficients $\mathfrak{C}^{\textsc{n},\textsc{b}}_n$ are the same as in \eqref{eq:Ecrv} and
\begin{equation}\label{eq:Bn}
	\begin{split}
	&\mathcal{B}_1=\frac14\int\limits_{0}^\infty\left[3\theta_0'(\rho)+\frac{\sin\theta_0(\rho)\cos\theta_0(\rho)}{\rho}\right]\rho^3\dd \rho,\\ &\mathcal{B}_2=\frac14\int\limits_{0}^\infty\sin^2\theta_0(\rho)\rho^3\dd \rho
	\end{split}
\end{equation} 
with $\rho=r/\ell$.

\section{Curvature induced skyrmion deformation}\label{app:deform}
The aim of this section is to obtain equilibrium values of the deformation amplitudes and the corresponding corrections to the energies of the deformed skyrmions. In order to be able to compare the influence of different types of deformations in different regimes, we first consider radially-symmetrical and elliptical deformation separately and then we consider the case when both these deformations are present at the same time.

\subsection{Taking into account the radially symmetrical deformation}\label{sec:rs}
Possible curvature-induced radially symmetrical deformation by means of the following Ansatz
\begin{equation}\label{eq:Ansatz-rad}
	\theta=\theta_0(rs),\qquad \Phi=\chi+\varphi,
\end{equation}
where scaling factor $s>0$ controls the change of the out-of-surface skyrmion component, while helicity $\varphi$ describes the conjugated change of the in-surface component. Dynamics of the variables $s$ and $\varphi$ reflects dynamics of the breathing magnon mode bounded on the skyrmion \cite{Kravchuk18,Kravchuk19a}.

Under the same assumptions as for the rigid skyrmion model one obtains 
\begin{equation}\label{eq:Eex-Neel-rs}
	\begin{split}
	E_\textsc{x}^{\textsc{n},\textsc{b}}=8\pi AL\biggl\{&C_0+\frac{\kappa(X_1)}{s}C_1\cos\varphi-\frac{C_2}{2}\frac{\kappa^2(X_1)}{s^2}\\
	&+\mathcal{O}(\kappa''r_0^3)+F[\kappa]\biggr\}
	\end{split}
\end{equation}
for the exchange energies and
\begin{equation}\label{eq:Edmi-tot-rs}
	\begin{split}
	&E_\textsc{d}^{\textsc{n}}=8\pi DL\left\{-\frac{\kappa(X_1)}{s^2}C_2+\frac{\cos\varphi C_1}{s}+\mathcal{O}(\kappa''r_0^4)+\bar{F}[\kappa]\right\},\\
	&E_\textsc{d}^{\textsc{b}}=8\pi DL\frac{C_1\sin\varphi}{s}
	\end{split}
\end{equation}
for the DMI energies. The anisotropy energy reads $E_\textsc{a}=8\pi KL C_2/s^2$. The total normalized energies are
\begin{subequations}\label{eq:Etot-rs}
	\begin{align}
		\label{eq:Etot-rs-N}&\mathcal{E}^{\textsc{n}}\approx\mathcal{C}_0-\frac{\mathcal{C}_2}{s}\left[2\sigma\cos\varphi-\frac{1}{s}+\varkappa\left(\frac{2}{|d|}\cos\varphi+\frac{d}{s}\right)+\frac{\varkappa^2}{2s}\right],\\
		\label{eq:Etot-rs-B}	&\mathcal{E}^{\textsc{b}}\approx\mathcal{C}_0-\frac{\mathcal{C}_2}{s}\left[2\sigma\sin\varphi-\frac{1}{s}+\frac{2\varkappa}{|d|}\cos\varphi+\frac{\varkappa^2}{2s}\right].
	\end{align}
\end{subequations}
Here we used the virial relation \eqref{eq:c1c2}. 

\emph{N{\'e}el skyrmion}. First of all, one can make sure that for the case $\varkappa=0$ the energy expression \eqref{eq:Etot-rs-N} is minimized for $\cos\varphi=\sigma$ and $s=1$. This corresponds to the undeformed planar N{\'e}el skyrmion with energy $\mathcal{E}^{\textsc{n}}_{\text{pl}}=\mathcal{C}_0-\mathcal{C}_2$. For the nonzero curvature, energy \eqref{eq:Etot-rs-N} is minimized for
\begin{equation}\label{eq:phi-s-Neel}
	\cos\varphi=\sgn{d+\varkappa},\qquad s=|d|\frac{1-\varkappa d-\frac{\varkappa^2}{2}}{|d+\varkappa|}.
\end{equation}
The corresponding expressions for $\tilde{\varphi}^\ts{n}=\varphi-(\sigma-1)\pi/2$ and $\tilde{s}^\ts{n}=s-1$ are presented in \eqref{eq:deform-rs-N}.

Substituting \eqref{eq:phi-s-Neel} into \eqref{eq:Etot-rs-N} we obtain the equilibrium energy in form $\mathcal{E}^{\textsc{n}}=\mathcal{E}^{\textsc{n}}_{\text{pl}}+\mathcal{E}^{\textsc{n}}_{\text{crv}}$, where the curvature-induced correction is $\mathcal{E}^{\textsc{n}}_{\text{crv}}=\mathfrak{C}^{\textsc{n}}_1\varkappa+\mathfrak{C}^{\textsc{n}}_2\frac{\varkappa^2}{2}+\mathcal{O}(\varkappa^3)$ with the coefficients $\mathfrak{C}^{\textsc{n}}_1=-\mathcal{C}_2\left(2+d^2\right)/d$ and $\mathfrak{C}^{\textsc{n}}_2=-\mathcal{C}_2\left(1+\Delta c_{\text{rs}}^\ts{n}\right)$. Thus, the coefficient $\mathfrak{C}^{\textsc{n}}_1$ at the leading term is not affected by the radial deformation, while the coefficient at $\varkappa^2$ obtains the deformation-induced correction $\Delta c_{\text{rs}}^\ts{n}=2(d^{-1}+d)^2$. 
%\begin{equation}\label{eq:E-Neel}
%	\begin{split}
%		&\mathcal{E}^{\textsc{n}}_{\text{crv}}=\mathfrak{C}^{\textsc{n}}_1\varkappa+\mathfrak{C}^{\textsc{n}}_2\frac{\varkappa^2}{2}+\mathcal{O}(\varkappa^3),\\
%		&\mathfrak{C}^{\textsc{n}}_1=-\mathcal{C}_2\left(\frac{2}{d}+d\right),\qquad \mathfrak{C}^{\textsc{n}}_2=-\mathcal{C}_2\left(\frac{2}{d^2}+5+2d^2\right).
%	\end{split}
%\end{equation}
%Comparison with \eqref{eq:ser-N} enables us to conclude that the radially symmetrical deformation of the skyrmion does not affect the linear in curvature term, however it results in the energy correction of the order $\varkappa^2$.

\emph{Bloch skyrmion}. First of all, one can make sure that for the case $\varkappa=0$ the energy expression \eqref{eq:Etot-rs-B} is minimized for $\sin\varphi=\sigma$ and $s=1$. This corresponds to the undeformed planar Bloch skyrmion with energy $\mathcal{E}^{\textsc{b}}_{\text{pl}}=\mathcal{C}_0-\mathcal{C}_2$. For the nonzero curvature, energy \eqref{eq:Etot-rs-B} is minimized for
\begin{equation}\label{eq:phi-s-Bloch}
	\tan\varphi=\frac{d}{\varkappa},\qquad s=|d|\frac{1-\frac{\varkappa^2}{2}}{\sqrt{d^2+\varkappa^2}}.
\end{equation}
The corresponding expressions for $\tilde{\varphi}^\ts{b}=\varphi-\sigma\pi/2$ and $\tilde{s}^\ts{b}=s-1$ are presented in \eqref{eq:deform-rs-B}.

Substituting \eqref{eq:phi-s-Bloch} into \eqref{eq:Etot-rs-B} we obtain the equilibrium energy in form $\mathcal{E}^{\textsc{b}}=\mathcal{E}^{\textsc{b}}_{\text{pl}}+\mathcal{E}^{\textsc{b}}_{\text{crv}}$, where the curvature-induced correction is $\mathcal{E}^{\textsc{b}}_{\text{crv}}=\mathfrak{C}^{\textsc{b}}_2\frac{\varkappa^2}{2}+\mathcal{O}(\varkappa^4)$. Here the coefficient $\mathfrak{C}^{\textsc{b}}_2=-\mathcal{C}_2(1+\Delta c_{\text{rs}}^\ts{b})$ obtains the deformation-induced correction $\Delta c_{\text{rs}}^\ts{b}=2/d^2$.
%\begin{equation}\label{eq:E-Bloch}
%		\mathcal{E}^{\textsc{b}}_{\text{crv}}=\mathfrak{C}^{\textsc{b}}_2\frac{\varkappa^2}{2}+\mathcal{O}(\varkappa^4),\quad \mathfrak{C}^{\textsc{b}}_2=-\mathcal{C}_2\left(\frac{2}{d^2}+1\right).
%\end{equation}
It is important to note that the deviations of $s$ from 1 determined by \eqref{eq:phi-s-Bloch} result in the energy corrections $\propto\varkappa^4$. This means that the only in-plane deformation $\varphi$ of the skyrmion profile can be taken into account for an approach with the terms $\mathcal{O}(\varkappa^4)$ neglected. 

\subsection{Taking into account the elliptical deformation}\label{sec:ell}
In order to descripe purely elliptical deformation we use \eqref{eq:Ansatz-rigid} with $\tilde{s}=\tilde{\varphi}=0$.
\begin{widetext}
In this case we obtain the following expressions for the exchange energy
\begin{equation}\label{eq:Eex-ell}
	\begin{split}
		E_\textsc{x}^{\textsc{n}}=&8\pi AL\Biggl\{C_0+ C_1\kappa\sigma-C_2\frac{\kappa^2}{2}+\frac{f^2}{2}\int\limits_0^\infty\theta_0'^2r\dd r+\frac{g^2}{2}\int\limits_0^\infty\frac{\sin^2\theta_0}{r}\dd r+C_1\frac{\sigma\kappa}{2}(f-g)\cos\lambda+\mathcal{O}(f\kappa^2r_0^2,g\kappa^2r_0^2,\kappa''r_0^2)+F[\kappa] \Biggr\},\\
		E_\textsc{x}^{\textsc{b}}=&8\pi AL\left\{C_0-C_2\frac{\kappa^2}{2}+\frac{f^2}{2}\int\limits_0^\infty\theta_0'^2r\dd r+\frac{g^2}{2}\int\limits_0^\infty\frac{\sin^2\theta_0}{r}\dd r+C_1\frac{\sigma\kappa}{2}(f-g)\sin\lambda+\mathcal{O}(f\kappa^2r_0^2,g\kappa^2r_0^2,\kappa''r_0^2)+F[\kappa] \right\},
	\end{split}
\end{equation}
\end{widetext}
for the DMI energy
\begin{equation}\label{eq:Edmi-ell}
	\begin{split}
		E_\textsc{d}^{\textsc{n}}=&8\pi DL\biggl\{C_1\sigma+\frac{C_1\sigma}{4}\left(2f^2-g^2\right)-\sigma C_1 fg\\
		&-\kappa C_2+\mathcal{O}(f^3r_0,g^3r_0,\kappa''r_0^4)\biggr\},\\
		E_\textsc{d}^{\textsc{b}}=&8\pi DL\biggl\{C_1\sigma+\frac{C_1\sigma}{4}\left(2f^2-g^2\right)-\sigma C_1 fg\\
		&-\frac{C_2\kappa}{2}(f+g)\sin\lambda+\mathcal{O}(f^3r_0,g^3r_0,\kappa''r_0^4)\biggr\},
	\end{split}
\end{equation}

and for the anisotropy energy
\begin{equation}\label{eq:Ean-ell}
	E_\textsc{a}^{\textsc{n},\textsc{b}}=8\pi KLC_2\left\{1+\frac32f^2+\mathcal{O}(f^4)\right\}.
\end{equation}
Using the virial relation \eqref{eq:c1c2} we present the total normalized energy in the form 
\begin{subequations}\label{eq:Etot-ell}
	\begin{align}
		\label{eq:Etot-ell-N}\mathcal{E}^{\textsc{n}}&\approx\mathcal{E}^{\textsc{n}}_{\text{pl}}+\mathcal{C}_2\biggl\{\alpha_f\frac{f^2}{2}+\alpha_g\frac{g^2}{2}-\frac{\varkappa}{d}\cos\lambda\left(f-g\right)\\\nonumber
		&+2fg-\varkappa\left(d+\frac{2}{d}\right)-\frac{\varkappa^2}{2}\biggr\},\\
		\label{eq:Etot-ell-B}\mathcal{E}^{\textsc{b}}&\approx\mathcal{E}^{\textsc{b}}_{\text{pl}}+\mathcal{C}_2\biggl\{\alpha_f\frac{f^2}{2}+\alpha_g\frac{g^2}{2}+2fg-\frac{\varkappa^2}{2}\\\nonumber
		&+\varkappa d\sin\lambda\left[g\left(\frac{1}{d^2}-\frac12\right)-f\left(\frac{1}{d^2}+\frac12\right)\right]\biggr\},
	\end{align}
\end{subequations}
where $\alpha_f$ and $\alpha_g$ are introduced in the main text.

\emph{N{\'e}el skyrmion}. First of all, one can make sure that for the case $\varkappa=0$ the energy expression \eqref{eq:Etot-ell-N} is minimized for $f^\textsc{n}=0$ and $g^\textsc{n}=0$. This corresponds to the undeformed planar N{\'e}el skyrmion with energy $\mathcal{E}^{\textsc{n}}_{\text{pl}}=\mathcal{C}_0-\mathcal{C}_2$. For the nonzero curvature, energy \eqref{eq:Etot-ell-N} is minimized if $\cos\lambda^\textsc{n} = 1$ and $f^\textsc{n}$, $g^\textsc{n}$ coincide with \eqref{eq:cv-ell-N} 
\footnote{Note that the alternative solution $\cos\lambda^\textsc{n}=-1$, $f^\textsc{n}\to-f^\textsc{n}$, $g^\textsc{n}\to-g^\textsc{n}$ corresponds to the same  magnetization configuration.\label{ft:cos}}. The corresponding curvature-dependent correction to the equilibium energy coincides with \eqref{eq:ser-N}. Coefficient $\mathfrak{C}_1^\textsc{n}$ is the same as in \eqref{eq:ser-N} and coefficient $\mathfrak{C}^{\textsc{n}}_2=-\mathcal{C}_2(1+\Delta c_{\text{el}}^\ts{n})$ has the deformation-induced correction 
\begin{equation}\label{eq:C2-ell-N}
 \Delta c_{\text{el}}^\ts{n}= \frac{1}{d^2}\frac{\alpha_f+\alpha_g+4}{\alpha_f\alpha_g-4}.
\end{equation}

\emph{Bloch skyrmion}. First of all, one can make sure that for the case $\varkappa=0$ the energy expression \eqref{eq:Etot-ell-B} is minimized for $f^\textsc{b}=0$ and $g^\textsc{b}=0$. This corresponds to the undeformed planar Bloch skyrmion with energy $\mathcal{E}^{\textsc{b}}_{\text{pl}}=\mathcal{C}_0-\mathcal{C}_2$. For the nonzero curvature, energy \eqref{eq:Etot-ell-B} is minimized if $\sin\lambda=1$ and the expressions for $f^\textsc{b}$ and $g^\textsc{b}$ are the same as in \eqref{eq:cv-ell-B} 
%\begin{equation}\label{eq:cv-ell-B}
%	\sin\lambda^\textsc{b}=1,\quad g^\textsc{b}=\frac{\varkappa d}{\alpha_f\alpha_g-4}\left[\frac{\alpha_f}{2}\left(1-\frac{2}{d^2}\right)-1-\frac{2}{d^2}\right],\quad f^\textsc{b}=\frac{\varkappa d}{\alpha_f\alpha_g-4}\left[\frac{\alpha_g}{2}\left(1+\frac{2}{d^2}\right)-1+\frac{2}{d^2}\right].
%\end{equation}
\footnote{Note that the alternative solution $\sin\lambda^\textsc{b}=-1$, $f^\textsc{b}\to-f^\textsc{b}$, $g^\textsc{b}\to-g^\textsc{b}$ corresponds to the same  magnetization configuration.}. 
The corresponding equilibrium energy coincides with \eqref{eq:ser-B} where the coefficient $\mathfrak{C}^{\textsc{b}}_2=-\mathcal{C}_2(1+\Delta c_{\text{el}}^\ts{b})$ has the deformation-induced correction 
\begin{equation}\label{eq:C2-ell-B}
 \Delta c_{\text{el}}^\ts{b}=\frac{d^2\left(\dfrac{\mathcal{C}_0}{\mathcal{C}_2}-\dfrac12\right)+\alpha_g-\alpha_f+\dfrac{4}{d^2}\left(\dfrac{\mathcal{C}_0}{\mathcal{C}_2}+\dfrac32\right)}{\alpha_f\alpha_g-4}.
\end{equation}
Here we used that $\alpha_f+\alpha_g=2+4\frac{\mathcal{C}_0}{\mathcal{C}_2}$.

\begin{figure}
	\includegraphics[width=0.8\columnwidth]{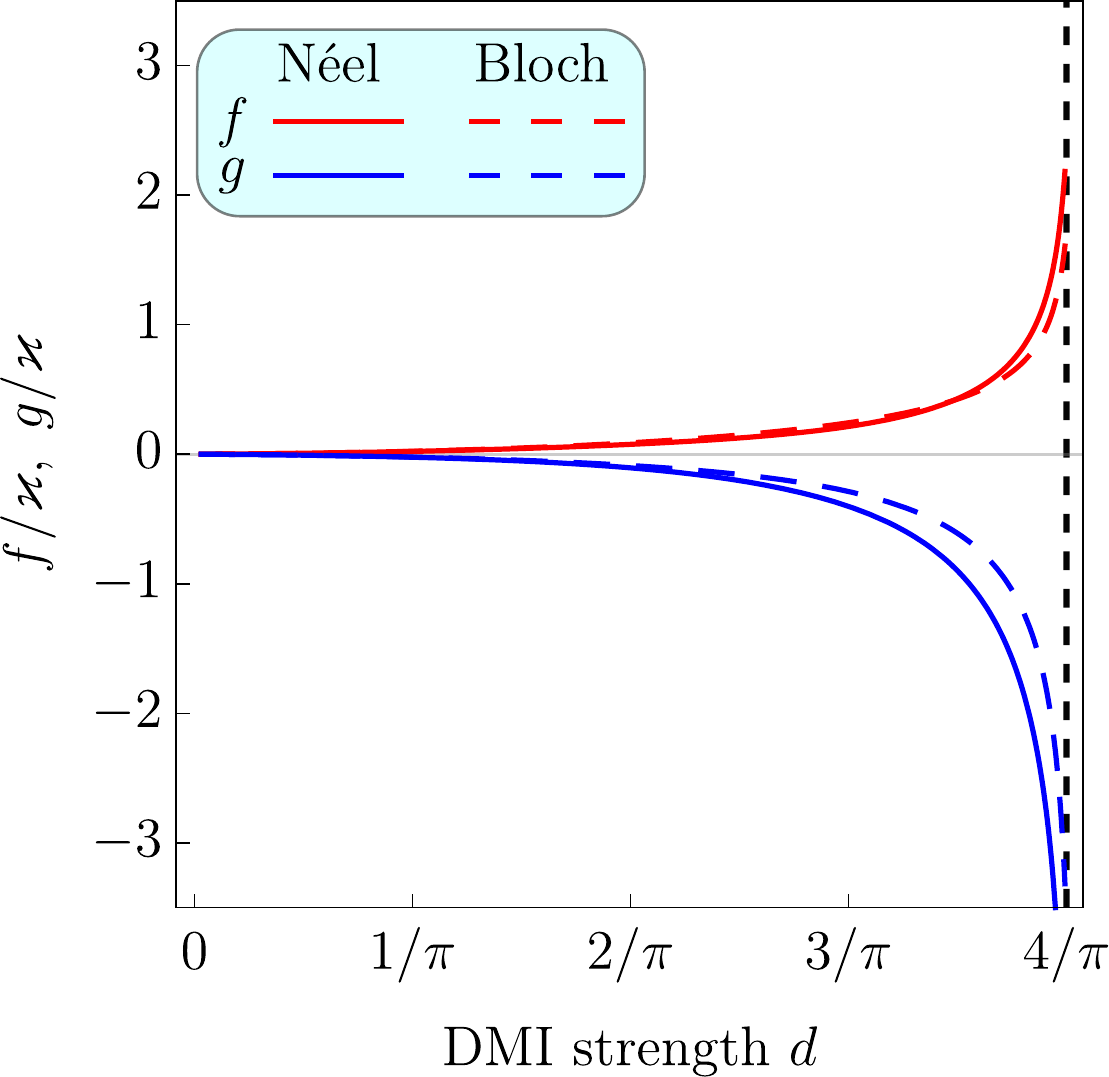}\caption{Dependence of amplitudes of elliptical deformation on DMI strength.}\label{fig:fg-vs-d}
\end{figure}

For both types of skyrmions, the dependencies of the deformation amplitudes on DMI constant is shown in Fig.~\ref{fig:fg-vs-d}.

\subsection{Taking into account the combined elliptical and radial symmetrical deformations}\label{sec:rs-ell-deform}
Here we consider the combined elliptical and radial-symmetrical deformation of N{\'e}el and Bloch skyrmions described by Ansatz \eqref{eq:Ansatz-ell-rad}. Performing he same steps as in the previous two subsections we obtain the following expression for the total normalized energies 
\begin{widetext}
%	In this case we obtain the following expressions for the exchange energy
%\begin{equation}\label{eq:Eex-ell-rad}
%	\begin{split}
%		E_\textsc{x}^{\textsc{n}}&\approx8\pi AL\biggl\{C_0+\frac{f^2}{2}\int\limits_0^\infty\theta_0'^2r\dd r+\frac{g^2}{2}\int\limits_0^\infty\frac{\sin^2\theta_0}{r}\dd r+\sigma\kappa C_1(1-\tilde{s})\cos\tilde{\varphi}+\frac{\sigma\kappa}{2}C_1(f-g)\cos(\lambda-\tilde{\varphi})-\frac{\kappa^2}{2}C_2+F[\kappa] \biggr\},\\
%		E_\textsc{x}^{\textsc{b}}&\approx8\pi AL\biggl\{C_0+\frac{f^2}{2}\int\limits_0^\infty\theta_0'^2r\dd r+\frac{g^2}{2}\int\limits_0^\infty\frac{\sin^2\theta_0}{r}\dd r-\sigma\kappa C_1\left(1-\tilde{s}\right)\sin\tilde{\varphi} +\frac{\sigma\kappa}{2}C_1(f-g)\sin(\lambda-\tilde{\varphi})-C_2\frac{\kappa^2}{2}+F[\kappa] \biggr\},
%	\end{split}
%\end{equation}
%for the DMI energy
%\begin{equation}\label{eq:Edmi-ell-rad}
%	\begin{split}
%		E_\textsc{d}^{\textsc{n}}&\approx8\pi DL\left\{\sigma C_1\left[1-\tilde{s}+\tilde{s}^2-\frac{g^2}{4}-fg+\frac{f^2}{2}\right]\cos\tilde{\varphi}-\kappa C_2\left(1-2\tilde{s}\right)\right\},\\
%		E_\textsc{d}^{\textsc{b}}&\approx8\pi DL\left\{\sigma C_1\left[1-\tilde{s}+\tilde{s}^2-\frac{g^2}{4}-fg+\frac{f^2}{2}\right]\cos\tilde{\varphi}-\frac{C_2\kappa}{2}(f+g)\sin\left(\lambda-2\tilde{\varphi}\right)\right\},
%	\end{split}
%\end{equation}
%and for the anisotropy energy $E_\textsc{a}^{\textsc{n},\textsc{b}}\approx8\pi KLC_2\left\{1-2\tilde{s}+3\tilde{s}^2+\frac32f^2\right\}$. Using the virial relation \eqref{eq:c1c2} we present the total normalized energy in the form 
\begin{subequations}\label{eq:Etot-ell-rad}
	\begin{align}
		\label{eq:Etot-ell-rad-N}
		\begin{split}
			\mathcal{E}^{\textsc{n}}\approx\mathcal{E}^{\textsc{n}}_{\text{pl}}+\mathcal{C}_2\biggl\{\alpha_f\frac{f^2}{2}+\alpha_g\frac{g^2}{2}&-\frac{\varkappa}{d}\left(f-g\right)\cos\left(\lambda-\tilde{\varphi}\right)+\left(2+f^2-\frac{g^2}{2}-2\tilde{s}+3\tilde{s}^2\right)\left(1-\cos\tilde{\varphi}\right)\\
			&+\left(2fg+\tilde{s}^2\right)\cos\tilde{\varphi}-d\varkappa\left(1-2\tilde{s}\right)-\frac{2\varkappa}{d}\left(1-\tilde{s}\right)\cos\tilde{\varphi}-\frac{\varkappa^2}{2}\biggr\},
		\end{split}\\
		\label{eq:Etot-ell-rad-B}
		\begin{split}
			\mathcal{E}^{\textsc{b}}\approx\mathcal{E}^{\textsc{b}}_{\text{pl}}+\mathcal{C}_2\biggl\{\alpha_f\frac{f^2}{2}+\alpha_g\frac{g^2}{2}&-\frac{\varkappa}{d}\left(f-g\right)\sin\left(\lambda-\tilde{\varphi}\right)+\left(2+f^2-\frac{g^2}{2}-2\tilde{s}+3\tilde{s}^2\right)\left(1-\cos\tilde{\varphi}\right)\\
			&+\left(2fg+\tilde{s}^2\right)\cos\tilde{\varphi}-\frac{d \varkappa}{2}\left(f+g\right)\sin\left(\lambda-2\tilde{\varphi}\right)+\frac{2\varkappa}{d}\left(1-\tilde{s}\right)\sin\tilde{\varphi}-\frac{\varkappa^2}{2}\biggr\},
		\end{split}
	\end{align}
\end{subequations}
where we assumed that the shape parameters $f$, $g$, $\tilde{s}$ and $\varkappa$ are of the same order of smallness and keep terms up to the second order.
\end{widetext}

\emph{N{\'e}el skyrmion}. First, note that energy \eqref{eq:Etot-ell-rad-N} coincides with energy~\eqref{eq:Etot-ell-N} of the elliptically deformed N{\'e}el skyrmion for $\tilde{\varphi}=0$ and $\tilde{s}=0$, and it coincides with energy \eqref{eq:Etot-rs-N} of radially symmetric N{\'e}el skyrmion if we substitute $f=g=0$, $\tilde{\varphi}=\varphi-\left(1-\sigma\right)\pi/2$ in \eqref{eq:Etot-ell-rad-N}  and $s=1+\tilde{s}$ in Eq.~\eqref{eq:Etot-rs-N} and keep the terms up to the second order in $\tilde{s}$. Assuming that $\tilde{\varphi}$ is of the same order of smallness as the other shape parameters we minimize energy \eqref{eq:Etot-ell-rad-N} for $\cos\lambda^\textsc{n}=1$, $\tilde{\varphi}^\textsc{n}\approx0$, $\tilde{s}^\textsc{n}=-\varkappa\left(d+d^{-1}\right)$ and $f^\textsc{n}$, $g^\textsc{n}$ defined in \eqref{eq:cv-ell-N} \cite{Note4}.
For the N{\'e}el skyrmion the tangential component of the deformation $\tilde{\varphi}$ is negligible while the normal component $\tilde{s}$ is not. 
%Note that the alternative solution $\cos\lambda^\textsc{n}=-1$, $f^\textsc{n}\to-f^\textsc{n}$, $g^\textsc{n}\to-g^\textsc{n}$, $\tilde{s}^\textsc{n}\to\tilde{s}^\textsc{n}$ corresponds to the same  magnetization configuration.  
The corresponding curvature-dependent correction to the equilibium energy coincides with \eqref{eq:ser-N}. Coefficient $\mathfrak{C}_1^\textsc{n}$ is the same as in \eqref{eq:ser-N} and coefficient $\mathfrak{C}_2^\textsc{n}=-\mathcal{C}_2(1+\Delta c^\ts{n})$ obtains the deformations induced correction $\Delta c^\ts{n}=\Delta c^\ts{n}_{\text{rs}}+\Delta c^\ts{n}_{\text{el}}$. 
%\begin{equation}\label{eq:E-Neel-eq}
% \mathfrak{C}^{\textsc{n}}_2=-\mathcal{C}_2\left(2d^2+5+\frac{2}{d^2}+\frac{1}{d^2}\frac{\alpha_f+\alpha_g+4}{\alpha_f\alpha_g-4}\right).
%\end{equation}

We conclude that the radial-symmetrical as well as elliptical deformations result in the energy corrections of order $\mathcal{O}(\varkappa^2)$, while the leading term remains independent on the deformations.

\emph{Bloch skyrmion}. First, note that energy \eqref{eq:Etot-ell-rad-B} coincides with energy~\eqref{eq:Etot-ell-B} of the elliptically deformed Bloch skyrmion for $\tilde{\varphi}=0$ and $\tilde{s}=0$, and it also coincides with energy \eqref{eq:Etot-rs-B} of radially symmetric Bloch skyrmion if we substitute $f=g=0$, $\tilde{\varphi}=\varphi-\sigma\pi/2$ in \eqref{eq:Etot-ell-rad-B} and  $s=1+\tilde{s}$ in \eqref{eq:Etot-rs-B} and keep terms up to the second order in $\tilde{s}$.  Assuming that $\tilde{\varphi}$ is of the same order of smallness as the other shape parameters we minimize energy energy \eqref{eq:Etot-ell-rad-B} for $\sin\lambda^\textsc{b}\approx1$, $\tilde{s}^\textsc{b}\approx0$, $\tilde{\varphi}^\textsc{b} \approx - \varkappa / d$ and expressions for $f^\textsc{b}$, $g^\textsc{b}$ coincides with \eqref{eq:cv-ell-B} \cite{Note5}. Here we assumed that $|\varkappa/d|\ll1$.  The corresponding curvature-dependent correction to the equilibium energy coincides with \eqref{eq:ser-B} with the coefficient $\mathfrak{C}^{\textsc{b}}_2=-\mathcal{C}_2(1+\Delta c^\ts{b})$, where $\Delta c^\ts{b}=\Delta c^\ts{b}_{\text{rs}}+\Delta c^\ts{b}_{\text{el}}$.
%\begin{equation}\label{eq:C2-ell-rad-B}
% \mathfrak{C}_2^{\textsc{b}}=-\mathcal{C}_2\left[1+\frac{\left(\dfrac{\mathcal{C}_0}{\mathcal{C}_2}-\dfrac{1}{2}\right)\left(\dfrac{4}{d^2}+d^2\right)+\alpha_g-\alpha_f+2\dfrac{\alpha_f\alpha_g}{d^2}}{\alpha_f\alpha_g-4}\right].
%\end{equation}

%For both types of skyrmions we conclude that the deformations of different types are independent if the derormation amplitudes are found in approximation linear in $\varkappa$. I.e. the elliptical deformation is not influenced by the radially-symmetrical one and vice-versa; the corresponding corrections $\Delta c^{\ts{n},\ts{b}}$ to the energy coefficients are additive.

\subsection{Comparison of different models}\label{sec:def-comparison}
\begin{figure*}
	\includegraphics[width=0.7\textwidth]{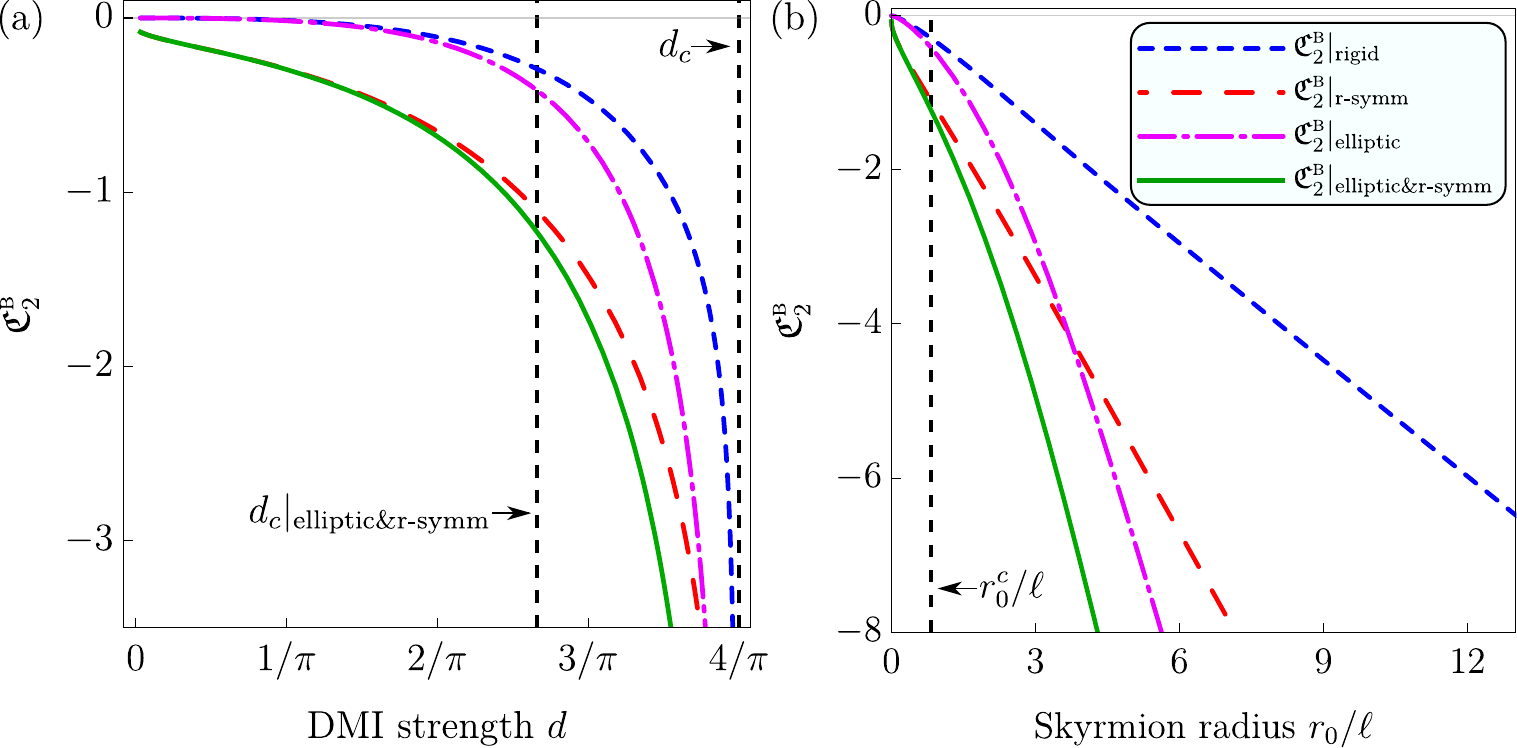}\caption{Influence of the skyrmion deformation on coefficient $\mathfrak{C}_2^{\textsc{b}}$. Blue, red, purple, and green lines correspond to the rigid skyrmion, radially-symmetrical deformation, elliptical deformation, and combined elliptical and radially-symmetrical deformations, respectively. Here $d_c= 4/\pi$ is a critical DMI value for planar systems.}\label{fig:CC}
\end{figure*}

As it was shown in the previous sections the curvature-induced energy corrections are linear $\mathcal{E}^\textsc{n}\approx\mathcal{E}^\textsc{n}_{\text{pl}}+\mathfrak{C}_1^{\textsc{n}}\varkappa$ and quadratic $\mathcal{E}^\textsc{b}\approx\mathcal{E}^\textsc{b}_{\text{pl}}+\mathfrak{C}_2^{\textsc{b}}\frac{\varkappa^2}{2}$ in curvature for N{\'e}el and Bloch skyrmion, respectively. Deformations of the skyrmion shape lead to the energy corrections which are quadratic in curvature for both types of skyrmions. This means that the effects of the shape deformation can be neglected for N{\'e}el skyrmions in the small curvature limit. This is in contrast to Bloch skyrmions for which the shape deformations must be taken into account. In order to determine which type of deformation is the most important one for the given Bloch skyrmion we compare the constants $\mathfrak{C}_2^{\textsc{b}}$ obtained for different deformation types, see Fig.~\ref{fig:CC}.
Both deformation types reduce skyrmion energy as compared to the rigid skyrmion. However, the relative impact of different deformations significantly depends on skyrmion radius $r_0$, which is unambiguously determined by the value of $d$ \cite{Kravchuk18,Komineas20a}.  From Fig.~\ref{fig:CC} one can conclude that the radially symmetrical deformation is dominant for small values of DMI strength $d<d_c\vert_\text{elliptic\&r-symm}\approx 0.85$ ($r_0<r_0^c\approx0.83\ell$), while for larger DMI values the combined elliptical and radially-symmetrical deformations must be taken into account. The value of $d_c\vert_\text{elliptic\&r-symm}$ was found as solution of equation $1-\mathfrak{C}_2^{\textsc{b}}\vert_\text{r-symm}[d]/\mathfrak{C}_2^{\textsc{b}}\vert_\text{elliptic\&r-symm}[d]=10^{-1}$.

%Note that the critical skyrmion radius $r_0^c$ is rather small, since for most magnets $\ell=3-10$ nm. This means that for the realistic cases the combined radially symmetric and elliptic deformations has to be taken into account for Bloch skyrmions.

\section{Dynamics of FM skyrmions}\label{app:dyn-fm}
Substitution of TWA into the Eq.~\eqref{eq:LLG} with the subsequent integration over the space domain results in the Thiele equation for collective coordinates \cite{Thiele73,Korniienko20}
\begin{equation}\label{eq:Thiele-tensor}
	\left(G_{\alpha\beta}-\eta_\ts{g}D_{\alpha\beta}\right)\partial_tX^\beta = \frac{\partial E}{\partial X^\alpha}.
\end{equation}
For details see Ref.~\onlinecite{Korniienko20}. Here $G_{\alpha\beta}=\varepsilon_{\alpha\beta}G$ is antisymmetric gyrotensor, whose magnitude $G=4\pi L\frac{M_s}{\gamma_0}N_{\text{top}}$ is proportional to the skyrmion topological charge $N_{\text{top}}$. Due to the Euclidean metric, the topological charge has the common form \cite{Korniienko20}, which in the angular parametrization reads $N_{\text{top}}=\frac{1}{4\pi}\int[\vec{\nabla}\theta\times\vec{\nabla}\phi]\vec{n}\,\dd x_1\dd x_2$. %of skyrmion in the limit of vanishing curvature, i.e.  $N_{\text{top}}=\frac{1}{4\pi}\int[\vec{\nabla}\theta_0\times\vec{\nabla}\phi_0]\vec{n}\,\dd x_1\dd x_2$, 
%In the following we assume that $N_{\text{top}}=-1$, this corresponds to the convention about the boundary conditions $\theta(0)=\pi$, $\theta(\infty)=0$ used in the previous section. 

The dissipation tensor is $D_{\alpha\beta}=LM_s\gamma_0^{-1}\mathbb{D}_{\alpha\beta}$, where
\begin{equation}\label{eq:Dab}
	\mathbb{D}_{\alpha\beta}=\iint\left[\partial_\alpha\theta\partial_\beta\theta+\sin^2\theta\partial_\alpha\phi\partial_\beta\phi\right]\dd x^1\dd x^2.
\end{equation} 
Note that for a constant metric tensor the expressions for $G_{\alpha\beta}$ and $D_{\alpha\beta}$ are exact consequences of TWA, i.e. the small curvature assumption is not required \cite{Korniienko20}. For N{\'e}el and Bloch skyrmions the dissipation tensors are $\mathbb{D}_{\alpha\beta}^{\textsc{n},\ts{b}}=4\pi\mathcal{C}_0\mathcal{D}_{\alpha\beta}^{\textsc{n},\ts{b}}$, where tensors $\mathcal{D}_{\alpha\beta}^{\textsc{n},\ts{b}}$ are defined in \eqref{eq:D-nb}. Introducing dimensionless time and distance (explained in Sec.~\ref{sec:dynamics-FM}) we write Eq.~\eqref{eq:Thiele-tensor} in form \eqref{eq:Thiele-tensor-dmnls}.

\underline{For the cylindrical Euler spiral}, energy of \emph{N{\'e}el} skyrmion \eqref{eq:ser-N} can be approximated as a linear function of the directix arc length: $\mathcal{E}^\ts{n}\approx\mathfrak{C}_1^\ts{n}\varkappa+\text{const}=\mathfrak{C}_1^\ts{n}k\x_1+\text{const}$. Substitution this energy into \eqref{eq:Thiele-tensor-dmnls} results in the equation of  motion which in the low-damping limit have solution 
\begin{equation}\label{eq:fm_neel_cornu_tr}
	\x_1(\tau)\approx\x_1(0)-\frac{\mathfrak{C}_1^{\ts{n}}k\eta}{N_{\text{top}}^2}\tau,\quad \x_2(\tau)\approx\x_2(0)+\frac{\mathfrak{C}_1^{\ts{n}}k}{N_{\text{top}}}\tau.
\end{equation}
Here the terms $\eta^2$ and $\eta\varkappa$ were neglected.

Energy of a \emph{Bloch} skyrmion is approximated as $\mathcal{E}^\ts{b}\approx\mathfrak{C}_2^{\ts{b}}\frac{\varkappa^2}{2}+\text{const}=-\frac12|\mathfrak{C}_2^{\ts{b}}|k^2\x_1^2+\text{const}$. This results into the following solutions of the equations of motion \eqref{eq:Thiele-tensor-dmnls}:
\begin{equation}\label{eq:fm_bloch_cornu_tr}
	\begin{split}
	&\x_1 \approx \x_1(0)\text{exp}\left(\dfrac{\eta|\mathfrak{C}_2^\ts{b}|\, k^2}{N_{\text{top}}^2}\tau\right),\\ 
	&\x_2 \approx \x_2(0)+\frac{\x_1(0)N_{\text{top}}}{\eta}\left[1-\exp\left(\dfrac{\eta|\mathfrak{C}_2^\ts{b}|\, k^2}{N_{\text{top}}^2}\tau\right)\right].
	\end{split}
\end{equation}
Absolute value of skyrmion velocity is
\begin{equation}\label{eq:fm_bloch_cornu_vel}
	|\vec{V}_\ts{fm}^\ts{b}| \approx \frac{|\x_1(0)\mathfrak{C}^\ts{b}_2|\, k^2}{|N_{\text{top}}|}\exp\left(\dfrac{\eta_\ts{fm}|\mathfrak{C}_2^\ts{b}|\, k^2}{N_{\text{top}}^2}\tau\right).
\end{equation}
\section{Dynamics of AFM skyrmions}\label{app:dyn-afm}
Equation \eqref{eq:n} can be obtained as the Euler-Lagrange equation for the Lagrange function
\begin{equation}\label{eq:L-afm}
	\mathcal{L}=\frac{M_s}{\gamma_0^2B_\ts{x}}\int\dot{\vec{\Omega}}^2\dd\vec{r}-E
\end{equation}
and Rayleigh dissipation function
\begin{equation}\label{eq:R-afm}
	\mathcal{R}=\eta_\ts{g}\frac{M_s}{\gamma_0}\int\dot{\vec{\Omega}}^2\dd\vec{r}
\end{equation}
with the constraint $|\vec{\Omega}|=1$ taken into account.

Substitution of TWA in \eqref{eq:L-afm} and \eqref{eq:R-afm} and integration over the spatial coordinates result in the following effective Lagrange and Rayleigh functions
\begin{equation}\label{eq:L-R-afm}
	\begin{split}
	&\mathcal{L}^{\text{eff}}=\frac{M_s}{\gamma_0^2B_\ts{x}}L\mathbb{D}_{\alpha\beta}\dot{X}^\alpha\dot{X}^\beta-E,\\ &\mathcal{R}^{\text{eff}}=\eta_\ts{g}\frac{M_s}{\gamma_0}L\mathbb{D}_{\alpha\beta}\dot{X}^\alpha\dot{X}^\beta.
	\end{split}
\end{equation}
Equations of motion generated by \eqref{eq:L-R-afm} and written in the dimensionless units have form \eqref{eq:eqs-afm}.

\underline{Cylindrical Euler spiral.} In the leading in curvature approximation, energy of a \emph{N{\'e}el} skyrmion is $\mathcal{E}^\ts{n}\approx\mathfrak{C}_1^\ts{n}\varkappa+\text{const}$, where $\varkappa(\x_1)=k\x_1$. In this case Eqs.~\eqref{eq:eqs-afm-N} have solution
\begin{equation}\label{eq:X1-sol}
	\begin{split}
	&\x_1(\bar\tau)=\x_1(0)+V_1(0)\frac{1-e^{-\bar\eta\bar\tau}}{\bar\eta}-a^{\textsc{n}}\frac{1-\bar\eta\bar\tau-e^{-\bar\eta\bar\tau}}{\bar\eta^2},\\	&\x_2(\bar\tau)=\x_2(0)+V_2(0)\frac{1-e^{-\bar\eta\bar\tau}}{\bar\eta}
	\end{split}
\end{equation}
where $V_\alpha=\dot{\x}_\alpha$ and $a^{\textsc{n}}=-\mathfrak{C}^\ts{n}_1k/\mathcal{C}_0$. 
%We introduce the notation $\tilde{\mathcal{D}}(\x^1)=\delta\varkappa(\x^1)$ based on the fact that $\tilde{\mathcal{D}}$ is linear in curvature, see Eq.~\eqref{eq:cv-ell-rad}. 
For the time intervals $\bar\tau\ll1/\bar\eta$ skyrmion demonstrates the uniformly accelerated dynamics
\begin{equation}\label{eq:X1-X2-Neel}
	\x_1(\bar\tau)\approx\x_1(0)+V_1(0)\bar\tau+\frac{a^{\textsc{n}}\bar\tau^2}{2},\; \x_2(\bar\tau)\approx\x_2(0)+V_2(0)\bar\tau,
\end{equation}
while in the long time limit $\bar\tau\gg1/\bar\eta$ skyrmion reaches a regime with constant velocity $\vec{V}_\ts{afm}^\ts{n}\approx\vec{e}_1a^{\textsc{n}}/\bar\eta$, see Fig.~\ref{fig:coord_AFM}(c). 

For a \emph{Bloch} skyrmion we estimate $\mathcal{E}^\ts{b}=-\frac{1}{2}|\mathfrak{C}_2^\ts{b}|\varkappa^2+\text{const}$. For this case  Eqs.~\eqref{eq:X1} has the solution 
\begin{equation}\label{eq:X2-sol-B}
		\x_1(\bar\tau)=\x_1(0)\biggl[\cosh(\Xi\bar\tau)+\frac{\bar\eta}{2\Xi}\sinh(\Xi\bar\tau)\biggr]e^{-\frac{\bar\eta}{2}\bar\tau},
\end{equation}
where $\Xi=\sqrt{k^2|\mathfrak{C}_2^\ts{b}|/\mathcal{C}_0+(\bar\eta/2)^2}$ and we assume the vanishing initial velocity. Note that for the vanishing curvature gradient $k=0$, Eq.~\eqref{eq:X2-sol-B} results in $\x_1(\bar\tau)=\x_1(0)$, i.e. the skyrmion is immobile. In the limit of small curvature gradients, solution \eqref{eq:X2-sol-B} has form 
\begin{equation}\label{eq:X1-ksmall}
	\x_1(\bar\tau)\approx\x_1(0)\left[1+k^2\frac{|\mathfrak{C}_2^\ts{b}|}{\mathcal{C}_0}\frac{e^{-\bar\eta\bar\tau}+\bar\eta\bar\tau-1}{\bar{\eta}^2}\right]
\end{equation}

Thus, in the initial time moments $\tau\ll1/\bar\eta$, skyrmion demonstrates a uniformly accelerated motion in $\vec{e}_1$ direction: $\x_1(\bar\tau)\approx\x_1(0)+a^{\textsc{b}}\bar\tau^2/2$, where $a^{\textsc{b}}=\x_1(0)k^2|\mathfrak{C}_2^\ts{b}|/\mathcal{C}_0$. In the opposite limit of long times $\bar\tau\gg1/\bar{\eta}$, the velocity component in $\x_1$ direction reaches constant value $\dot{\x}_1\approx a^{\textsc{b}}/\bar{\eta}$.

Using the fact that $\delta^\ts{b}$ is linear function of the curvature, we present Eq.~\eqref{eq:X2} in form
\begin{equation}\label{eq:X2-afm}
	\ddot{\x}_2+\bar{\eta}\dot{\x}_2=\frac{\Delta^\ts{b}k^3|\mathfrak{C}_2^\ts{b}|}{\mathcal{C}_0}\x_1^2(\bar\tau),
\end{equation}
where $\Delta^\ts{b}=\delta^\ts{b}/\varkappa$ (this quantity is shown in Fig.~\ref{fig:D}). Substituting \eqref{eq:X2-sol-B} into \eqref{eq:X2-afm} we obtain a solution, which in the limit of small gradients has the form
\begin{equation}\label{eq:X2-ksmall}
	\x_2(\bar\tau)\approx\x_2(0)+\tilde{a}^\ts{b}\frac{e^{-\bar\eta\bar\tau}+\bar\eta\bar\tau-1}{\bar{\eta}^2}
\end{equation}
with $\tilde{a}^\ts{b}=\x_1(0)k\Delta^\ts{b}a^\ts{b}$. Solving \eqref{eq:X2-ksmall} we assumed vanishing initial velocity. Thus, we have the uniformly accelerated motion in the initial moments of time and the reaching of constant velocity $\dot{\x}_2\approx\tilde{a}^\ts{b}/\bar{\eta}$ for $\bar\tau\gg1/\bar{\eta}$. Finally we conclude that in the long time limit Bloch skyrmion reaches a constant velocity $\vec{V}_\ts{afm}^\ts{b}\approx\left(\vec{e}_1a^{\textsc{b}}+\vec{e}_2\tilde{a}^{\textsc{b}}\right)/\bar\eta$.

\section{Numerical simulations}\label{app:simuls}

\subsection{Spin-lattice simulations}\label{app:sl-sim}

In order to verify our analytical calculations we perform a set numerical simulations for FM/AFM curved films. We consider the generalized cylindrical surface as a square lattice with
lattice constant $a$. Each node is characterized by a magnetic moment $\vec{m}_{\vec{p}}(t)$ which is located at the position $\vec{r}_{\vec{p}}$. Here $\vec{p}=(i,j)\in\mathbb{N}\times\mathbb{N}$ defines the magnetic moment and its position on the lattice with size $N_1\times N_2$, i.e. $i\in\left[1,N_1\right]$ and $j\in\left[1,N_2\right]$. The dynamics of magnetic system is govern by discrete Landau--Lifshitz--Gilbert equations 
\begin{equation}\label{eq:llg_sim}
	\frac{\mathrm{d}\vec{m}_{\vec{p}}}{\mathrm{d}t} = \frac{\gamma_0}{\mu_s}\left[\vec{m}_{\vec{p}}\times\frac{\partial\mathscr{H}}{\partial\vec{m}_{\vec{p}}}\right]+\eta_\textsc{g}\,\left[\vec{m}_{\vec{p}}\times\frac{\mathrm{d}\vec{m}_{\vec{p}}}{\mathrm{d}t}\right],
\end{equation}
where $\mu_s$ is a magnetic moment of a magnetic site. The Hamiltonian of a magnetic system has following form
\begin{equation}\label{eq:hamiltonian}
	\begin{split}
		\mathscr{H} = - \frac{\mathscr{J}}{2} \sum \vec{m}_{\vec{p}} \cdot \vec{m}_{\overline{\vec{p}}} - \frac{\mathscr{K}}{2}\sum \left(\vec{m}_{\vec{p}}\cdot\vec{n}_{\vec{p}}\right)^2\\ 
		+ \frac{\mathscr{D}}{2} \sum \vec{d}_{\vec{p},\overline{\vec{p}}}\cdot\left[\vec{m}_{\vec{p}} \times \vec{m}_{\overline{\vec{p}}}\right].
	\end{split}
\end{equation}
Here $\mathscr{J}$ is an effective exchange integral ($\mathscr{J} > 0$ for FM and $\mathscr{J} < 0$ for AFM ordering), $\mathscr{D}$ is an effective DMI strength constant, $\mathscr{K}>0$ is an effective easy-normal anisottropy constant, $\overline{\vec{p}}$ runs over nearest neighbors, $\vec{d}_{\vec{p},\overline{\vec{p}}}$ is a DMI vector ($\vec{d}_{\vec{p},\overline{\vec{p}}} = \vec{n}_{\vec{p}}\times\vec{u}_{\vec{p},\overline{\vec{p}}}$ for the interfacial DM with $\vec{u}_{\vec{p},\overline{\vec{p}}} = \left(\vec{r}_{\vec{p}}-\vec{r}_{\overline{\vec{p}}}\right)/a$, and $\vec{d}_{\vec{p},\overline{\vec{p}}} = \vec{u}_{\vec{p},\overline{\vec{p}}}$ for isotropic DMI), and $\vec{n}_{\vec{p}}=\vec{n}(ai,aj)$ is a normal unit vector to the surface. Note that the normal $\vec{n}_{\vec{p}}$ introduces the information about the surface shape into the model \eqref{eq:hamiltonian}.

The length scale in simulations is defined with the magnetic length as $\ell = a \sqrt{|\mathscr{J}|/\mathscr{K}}$, the dimensionless DMI strength is $d = \mathscr{D}/\sqrt{\mathscr{|J|K}}$, the dimensionless time for FM system is $\tau = 2\gamma_0 \mathscr{K} t/\mu_s$, and the dimensionless time for the AFM system is $\overline{\tau} = 2\gamma_0 t \sqrt{2\mathscr{|J|K}}/\mu_s$.   

The magnetization dynamics is simulated by means of numerical solution of the set of ODE \eqref{eq:llg_sim} for the initial conditions determined by the initial magnetization.

\subsubsection{Simulations of sinusoidal-shaped film}
\underline{Skyrmions condensation.} We considered the sinusoidal surface with directrix $\vec{\gamma}$ defined with~\eqref{eq:sin}. In simulations we considered sinusoidal surface with $N_1 = 401$ and $N_2 = 201$, amplitude $\mathcal{A} = 10\ell$, period $\mathcal{T} = 35\ell$, DMI strength $d = 1$, Gilbert damping $\eta_\ts{g} = 0.1$, and magnetic length $\ell \approx 7.5a$. These parameters result in a one period along $\vec{e}_1$ direction. In order to simulate more than one period we impose the periodic boundary conditions along $\vec{e}_1$ and $\vec{e}_2$ directions.

We performed four different simulations: (i) FM film with N{\'e}el DMI; (ii) FM film with Bloch DMI; (iii) AFM film with N{\'e}el DMI; (iv) AFM film with Bloch DMI. All simulations for sinusoidal surface were performed in the same manner. Namely, we set initial state of a system with a hexagonal skyrmion lattice with the distance between the skyrmions $80a$. Afterwards, we simulate the relaxation dynamics, see Supplemental movies. Equilibrium states for the sinusoidal surfaces are presented in Fig.~\ref{fig:sine_geometry}.

\underline{Oscillation of AFM skyrmion.} We simulated sinusoidal surfaces with the period $\mathcal{T} = 35\ell$, DMI strength $d = 1$, Gilbert damping parameter $\eta_{\textsc{g}} = 10^{-4}$, and amplitude in range $\mathcal{A} \in \left[0.1,1\right]\ell$ with $\Delta\mathcal{A}=0.1\ell$. Here we simulate film with the dimensions $N_1 = 401$ and $N_2=201$ (here we did not impose periodic boundary conditions). The N{\'e}el and Bloch skyrmions were placed in position with $X_1(0) = X_1^0 + 2\ell$, where $X_1^0$ is the position of stable equilibrium. Afterward, we observed free dynamics of skyrmion, which corresponds to the decaying oscillations with well pronounced oscillations $\omega_p$, see Fig.~\ref{fig:oscillations}. The corresponding curvature-induced motion of the skyrmons are presented in the supplemental movies.

\subsubsection{Simulations of Euler spiral-shaped film}

We considered the Euler spiral-shaped film with directrix defined in~\eqref{eq:Euler_spiral}. In simulations we considered surfaces with $N_1 = 301$ and $N_2 = 501$, DMI strength $d = 1$, magnetic length $\ell = 5a$, and gradient of the curvature $k\in \left[2,10\right]\times 10^{-3}$ with $\Delta k = 2\times 10^{-3}$. 

Similarly to the case of sinusoidal geometry, here we also performed four sets of different simulations. The simulations were performed in two steps. Firstly, we relaxed magnetic skyrmion in an overdamped regime with $\eta_\ts{g} = 0.5$. The initial positions were set as follows:
\begin{itemize}
	\item FM N{\'e}el skyrmion: $\x_1(0) = 0$ and $\x_2(0) = -30$;
	\item FM Bloch skyrmion: $\x_1(0) = -15$ and $\x_2(0) = 0$;
	\item AFM N{\'e}el skyrmion: $\x_1(0) = \x_2(0) = 0$;
	\item AFM Bloch skyrmion: $\x_1(0) = -15$ and $\x_2(0) = 0$.
\end{itemize}
The initial position for N{\'e}el skyrmions corresponds to the mean curvature with $\varkappa[\x_1(0)] = 0$, while for the Bloch skyrmons we shifted skyrmions to the position with $\varkappa[\x_1(0)]\neq 0$ in order to avoid the unstable equilibrium.

On the second step, we observed free dynamics with Gilbert damping $\eta_\ts{g} = 0.1$. The corresponding curvature-induced motion of the skyrmons are presented in the supplemental movies. The trajectories of the skyrmions are presented in Figs.~\ref{fig:coord_FM} and \ref{fig:coord_AFM} for FM and AFM ordering, respectively. The center of skyrmions was defined as the first moment of the gyrovector density
%\begin{equation}
%	\mathcal{X}_\alpha = \frac{\int_\mathcal{S}\xi_\alpha\left(1-\Omega_n\right)\mathrm{d}\mathcal{S}}{\int_\mathcal{S}\left(1-\Omega_n\right)\mathrm{d}\mathcal{S}}.
%\end{equation}
\begin{equation}
	\mathcal{X}_\alpha = \frac{1}{4\pi N_{\text{top}}}\int_\mathcal{S}\xi_\alpha\mathfrak{g}\,\dd\mathcal{S},\quad \mathfrak{g}=\epsilon_{ijk}m_i\partial_{1}m_j\partial_2m_k.
\end{equation}
Alternatively one can write $\mathfrak{g}=[\vec{\nabla}\theta\times\vec{\nabla}\phi]\cdot\vec{n}$.

\subsubsection{Movies of the skyrmion motion}

For a better illustration of the skyrmion motion induced by the curvature gradients, we  prepared  movies  of skyrmion dynamics.  Movies are based on data obtained by means of the numerical simulations. 
\begin{itemize}
	\item \texttt{FM\_skyrmion\_latt\_d\_0.95\_A\_10ell\_T\_35ell.mp4} shows the dynamics of the FM N{\'e}el and Bloch skyrmions in the sinusoidal-shaped film with the amplitude $\mathcal{A} = 10\ell$, period $\mathcal{T} = 35\ell$, DMI strength $d = 0.95$, and Gilbert damping $\eta_\textsc{g} = 0.1$.
	\item \texttt{AFM\_skyrmions\_osc\_d\_1\_A\_1ell\_T\_35ell.mp4} shows the dynamics of the AFM Bloch skyrmions in the sinusoidal-shaped film with the amplitude $\mathcal{A} = 1\ell$, period $\mathcal{T} = 35\ell$, DMI strength $d = 1$, and Gilbert damping $\eta_\textsc{g} = 10^{-4}$.
	\item \texttt{FM\_skyrmions\_cornu\_spiral\_d\_1\_k\_6e-3\_eta\_0.1.mp4}  shows the dynamics of the FM N{\'e}el and Bloch skyrmions in the Euler spiral with the gradient of the curvature $k = 6\times 10^{-3}$, DMI strength $d = 1$, and Gilbert damping $\eta_\textsc{g} = 0.1$.
	\item \texttt{AFM\_skyrmions\_cornu\_spiral\_d\_1\_k\_6e-3\_eta\_0.1.mp4}  shows the dynamics of the AFM N{\'e}el and Bloch skyrmions in the Euler spiral with the gradient of the curvature $k = 6\times 10^{-3}$, DMI strength $d = 1$, and Gilbert damping $\eta_\textsc{g} = 0.1$.
\end{itemize}

\subsection{Full-scale micromagnetic simulations}\label{app:mm-sim}
In order to cross check our results we also performed a series of full-scale micromagnetic simulations using our GPU accelerated finite-element code, TetraMag~\cite{kakaySpeedupFEMMicromagnetic2010}. The static equilibrium states of the skyrmions on the sinusoidal-shaped films and Euler spiral-shaped films as well as their dynamics (curvature-induce drift) were obtained by the numerical integration of the Landau-Lifshitz-Gilbert equation. The following material parameters were considered in the simulations: $A = 1.6 \times 10^{11}$ J/m being the exchange constant, $\mu_0 M_s = 1.38$ T the saturation magnetization, $K_u = 1.3 \times 10^6$ J/$\text{m}^3$ for the uniaxial anisotropy constant pointing along the surface normal and $d = D/\sqrt{AK_{eff}}$ for the dimensionless DMI constant. The effective anisotropy constant for the renormalised magnetostatic case was set to $K_{eff} = K_u  - 2 \pi M_s^2 = 5.1 \times 10^5$ J/$\text{m}^3$. The magnetic length is therefore $\ell=\sqrt{A/K_{eff}} =5.6$ nm. The thickness of all studied films is set to \SI{1.5}{\nano\meter} and the samples discretized with an average tetrahedron edge length of \SI{1}{\nano\meter}.

%\bibliography{skyrmion_drift}

\end{document}